\documentstyle[12pt,aaspp4]{article}

\begin{document}

\title{A Mid-Infrared Galaxy Atlas (MIGA)} 

\author{C. R. Kerton\altaffilmark{1} and P. G. Martin}

\affil{Canadian Institute for Theoretical Astrophysics, University of
Toronto, Toronto, Ontario M5S~3H8, Canada}

\altaffiltext{1}{Also at Department of Astronomy, University of Toronto}

\authoremail{kerton, pgmartin@cita.utoronto.ca}

\begin{abstract}
A mid-infrared atlas of part of the Galactic plane ($75^\circ < l <
148^\circ, b = \pm6^\circ$) has been constructed using HIRES processed
infrared data to provide a mid-infrared data set for the Canadian
Galactic Plane Survey (CGPS). The addition of this data set to the
CGPS will enable the study of the emission from the smallest
components of interstellar dust at an angular resolution comparable to
that of the radio, millimetre, and far-infrared data in the CGPS. The
Mid-Infrared Galaxy Atlas (MIGA) is a mid-infrared (12 $\mu$m and 25
$\mu$m) counterpart to the far-infrared IRAS Galaxy Atlas (IGA), and
consists of resolution enhanced ($\sim 0.5'$ resolution) HIRES images
along with ancillary maps. This paper describes the processing and
characteristics of the atlas, the cross-beam simulation technique used
to obtain high-resolution ratio maps, and future plans to extend both the
IGA and MIGA.   

\end{abstract}


\section{Introduction}
\label{sec:intro}

The Mid-Infrared Galaxy Atlas (MIGA) is a mid-infrared (12 and
25~$\mu$m) atlas of part of the Galactic plane ($75^\circ < l <
148^\circ, b = \pm6^\circ$).  It was constructed
using IRAS data processed to approximately $0.5'$ resolution using the
HIRES image construction process (\cite{aum90}) including a new point
source ringing suppression algorithm (\cite{cao99}).   Parts
of the MIGA along with the far-infrared (60 and 100 $\mu$m)
IRAS Galaxy Atlas (IGA; \cite{cao97}) are being merged with
radio and  millimetre data as part of the Canadian Galactic Plane
Survey  (\cite{gps98}; CGPS\footnote{Current information on the CGPS
can be found at http://www.ras.ucalgary.ca/CGPS/}), a project to
survey about a quadrant of the Galactic plane  at $\leq 1'$
resolution over a wide range of wavelengths (12 $\mu$m --  190 cm) to
study all of the major components of the interstellar medium (ISM).

The addition of a mid-infrared data set to this data base is important
since  one of the goals of the CGPS is to understand the evolution
of dust as  it moves through different phases of the ISM. The 25 and
12 micron  bands of IRAS have been shown to be good tracers of
the smallest  dust particles: the chemically uncharacterized very small grains
(VSG's) and the large carbonaceous molecules, most likely polycyclic
aromatic hydrocarbons (PAH's) (\cite{ona96}), respectively.  Therefore, the
MIGA will enable the study of the emission from the smallest 
components of interstellar dust at an angular resolution 
comparable to that of the complementary data being used to define
different physical environments.

This paper has been designed to be complementary to the paper
describing the IGA (\cite{cao97}). Many of the details of the image
construction algorithms for the MIGA and the IGA are identical and were
described in great detail in a series of papers related to both the
IGA and the parallelization of the HIRES code (\cite{cao97};
\cite{cao96}); thus they will not be repeated here.  Rather we
concentrate on pointing out those areas where the MIGA and IGA differ
(most importantly in resolution behaviour and the response to point sources)
and on demonstrating through a series of tests on the data that the
MIGA is a data set of comparable quality to the IGA.  This paper, combined
with the papers describing the IGA,  gives a complete
guide to the infrared data sets that will be made available to the
general astronomical community as part of the CGPS data releases over
the next few years. The information provided in this paper is relevant
to both  MIGA mosaiced images included in the CGPS and stand-alone MIGA images.

In \S~\ref{sec:atlas} we describe the format of the atlas along
with the format of an extension to the IGA (EIGA) that was constructed
specifically for the CGPS.  The steps involved in MIGA processing are
outlined in \S~\ref{sec:processing}.  Section \ref{sec:character}
describes the characteristics of MIGA images.  In
\S~\ref{sec:artifacts} various artifacts of the images are discussed,
including the reduced point source ringing.  Finally, sample images are
shown in \S~\ref{sec:images} and future directions for the MIGA
and large-scale HIRES processing are discussed in \S~\ref{sec:sumfut}.

\section{Description of the Atlas --- MIGA and EIGA}
\label{sec:atlas}
The atlas covers a twelve degree wide strip ($b = \pm6^\circ$) of the
Galactic plane from Cygnus to Cassiopeia ($75^\circ < l <
148^\circ$). The higher latitude limit, compared to the
IGA ($b = \pm4.7^\circ$), was required to match the MIGA with the CGPS
coverage which extends from $ -3.56^\circ < b <  5.56^\circ$ in order to
follow the main concentration of HI in this part of the Galaxy.  With
this upper bound set, the lower boundary of the atlas was chosen to be
symmetric.

Each MIGA image covers a $1.4^\circ \times 1.4^\circ$ area and consists of
$1^{\mathrm st}$ and $20^{\mathrm th}$ iteration HIRES images along
with ten ancillary maps and tables as listed in Table~\ref{tbl:files}.
See Figures~\ref{fig:allimage_2} and~\ref{fig:allimage_1} for sample images.
The images are in Galactic cartesian (CAR) projection (\cite{gre96})
with a pixel size of $15''$.  MIGA images are quite flexible; users
can make seamless mosaics of arbitrary size (see \S~\ref{sec:images})
and rebin and reproject the images as required. We hope to make the
full MIGA available via the web in a similar manner to the existing IGA web
server\footnote{http://irsa.ipac.caltech.edu/applications/IGA/}. 

Part of the MIGA, a series of $5.12^\circ \times 5.12^\circ$
mosaics, will be available
through the Canadian Astronomy Data Centre (CADC) as part of the
general public release of the CGPS data.  The CGPS mosaics have a
pixel size of $18''$, and so the MIGA and IGA 
images were slightly rebinned in creating these mosaics.
The CGPS mosaics will be a very useful way to access both the MIGA and
IGA data, particularly since users will have immediate access to
complementary radio and millimetre data on a uniform grid covering the
same region.  Tests done on regular and rebinned MIGA images show that
there is little reduction in the quality of the images due to the
slight rebinning.  The general image characteristics described here
for the MIGA, and in Cao et al.\ (1997) for the IGA, apply equally to
the MIGA and IGA in the CGPS mosaic format.

Since the IGA has a high-latitude cutoff of $b = 4.7^\circ$ the
initial CGPS mosaics constructed using the IGA had a blank strip at
high latitudes.  In order to match the infrared coverage of the CGPS
with the radio coverage we constructed a high-latitude extension to
the IGA covering $75^\circ < l <148^\circ$
and $ 4.7^\circ < b < 5.56^\circ$.  Agreement between the original IGA
images and the EIGA is excellent and the EIGA has been incorporated
into all of the CGPS far-infrared mosaics (see \S~\ref{sec:sumfut}).
We have also constructed a low latitude extension to the IGA from $
-6.0^\circ < b < -4.7^\circ$ to match the IGA and MIGA coverage.
This ability to extend the infrared images of the IGA and the MIGA to
higher/lower Galactic latitudes is an important reason why HIRES and
the IRAS data base, with its almost full-sky coverage, remains an
important tool to study the infrared sky.  This will be discussed more
in \S~\ref{sec:sumfut}. 

\section{Description of Processing}
\label{sec:processing}

The basic processing of the MIGA (and EIGA) follows the same steps as
discussed in detail in Cao et al.\ (1997).  The raw IRAS archive data,
known as CRDD (Calibrated Reconstructed Detector Data) were first sent from
IPAC to the Canadian Institute for Theoretical Astrophysics (CITA) to
allow processing to be done locally. The raw data are
uncompressed and formatted using the programs SnipScan and LAUNDR.
The data are processed in $7^\circ \times 7^\circ$ sections known as
CRDD plates.

Infrared Sky Survey Atlas (ISSA) images corresponding to the plate region are
mosaiced together and used to calibrate the IRAS data using the SmLAUN
program.  This step effectively removes the zodiacal emission since
ISSA images have a zodiacal light model subtracted from them. ISSA
mosaicing was done in advance of the other preprocessing steps to
allow the quality of the mosaicing to be checked before use 
(see \S~\ref{sec:artifacts}).  The data are then reprojected from
equatorial coordinates to Galactic coordinates using the BrkDet
program.  At this stage the data are in $1.4^\circ \times 1.4^\circ$
sections spaced every $1^\circ$. All of the preprocessing steps were
done on a Sparc Ultra workstation. For a single CRDD plate
preprocessing took on average about two hours of wall clock time to
complete in addition to the time required to construct and check the
ISSA mosaics.  To cover the CGPS region 63 CRDD plates in two bands
had to be processed ($\sim240$ hours).

These data are then HIRES processed to create the final images and ancillary
maps. Unlike the IGA, we used a non-parallel version of the
HIRES code on a SGI Origin 2000 computer at CITA.  The
original HIRES code was first modified to run on the SGI architecture and
tests were made comparing IGA release images to 60 and 100 $\mu$m
images constructed at CITA.  IGA images were recreated for a field at
$l = 152^\circ, b = -1^\circ$ using the new code and no differences
were seen, beyond that expected for numerical round-off errors:
maximum fractional differences were on the order of $10^{-5}$ and average
fractional differences were on the order of $10^{-8}$.  

The construction of the EIGA gave us another chance to
test the new code for compatibility with the IGA production
code. As shown in \S~\ref{sec:sumfut} the match between the IGA images
and the EIGA images produced at CITA is excellent. 

Tests showed that processing a single $1.4^\circ \times 1.4^\circ$
region at a given wavelength 
took approximately 4.5 minutes of wall clock time.  While this is clearly
slower than the processing times reported in Cao et al.\ (1996) for the
parallel-processing machines, it is a vast improvement over the
single-processor times they report. Since we were primarily interested
in covering the CGPS survey region which contains 444 $1.4^\circ \times
1.4^\circ$ areas, and considering other overheads in the production,
this processing speed was adequate.  It took 67 hours of wall clock
time to process the entire CGPS region at both 12 and 25 $\mu$m (about a 
quarter of the time required to prepare the CRDD data on the Sparc Ultra).

The HIRES code we used also took advantage of a ringing suppression
algorithm that was developed after the IGA was in its production run 
(\cite{cao99}). Tests were done before producing the MIGA to compare
the quality of the images with and without the ringing suppression
algorithm in place.  Differences between the two images were negligible
away from point sources.  In \S~\ref{sec:rings} we discuss the
effect of the ringing suppression algorithm on point sources in more detail.

\section {Characteristics of the Images}
\label{sec:character}
In this section of the paper we describe the resolution, photometric
accuracy, positional accuracy, surface brightness accuracy, and the mosaic
property of the MIGA.

\subsection{Resolution}
\label{sec:resolution}

The basic angular size of the rectangular IRAS detectors is $45''
\times 270''$ at 12 $\mu$m and $45'' \times 276''$ at 25 $\mu$m
(\cite{aum90}).  The higher resolution is obtained along the scan
direction of the satellite. As a result of different scan orientations
the position angle of the elongated raw beam varies across the
sky. Improved resolution is possible because each region is covered by
overlapping scans and was usually revisited with a different scan 
orientation (see Figure~\ref{fig:allimage_1}).  

In order to assess the achieved resolution of the MIGA within each
field simulated beams (PSFs) are constructed using the HIRES IRAS
Simulator mode (\cite{fow94}). Spikes are placed in a regular grid on
a smoothed version of the 20$^{\mathrm th}$ image.  As discussed in
Cao et al.\ (1997), the image histogram is used to scale the spikes to an
intensity that represents a point source that is bright enough,
relative to the background emission, that HIRES processing is beneficial. 
The image is then scanned with the detector pattern to produce simulated
IRAS data.  These data are then regularly HIRES processed to produce
the img\_*bem* maps that show the IRAS beam shape (see
Figure~\ref{fig:allimage_2}).  As will be discussed below, it is
important to note that while the results of a 2-D Gaussian fit to
these beams does give a measure of the resolution (*fwhm.txt files),
the actual beam shapes are not 2-D Gaussians even when the beams are
not X-shaped due to large differences in the scan directions
(\cite{ric93}, \cite{mos92}).  

To quantify the MIGA resolution we sampled twelve fields scattered
across the atlas region, using the FWHM of 2-D Gaussian fits to
simulated beams as a measure of the achieved resolution.
The results of this test are shown in Table~\ref{tbl:restest} where we
report the mean and standard deviation about the mean for the 49 beams
in each field.  The average resolution for the twelve test fields is
$33'' \times 67''$ and $34'' \times 66''$ at 12 and 25 $\mu$m
respectively. This should be compared with the typical
full-resolution co-add (FRESCO; equivalent to 1$^{\mathrm st}$
iteration MIGA and IGA images) resolution of $1' \times 5'$ in both
wavebands or the standardized ISSA resolution of $4' \times 5'$ in all
wavebands. Note that the pixel size of 15$''$ just adequately samples
the beam in the scan direction.

One important thing to notice is the similarity of the resolution at
12 and 25 $\mu$m.  Within any given field the resolution may vary
from place to place, but the 12 and 25 $\mu$m resolutions are always
very close in value.  It is also important to note that the position
angles of the beams are also very close;  in a test of 756 simulated
12 and 25 $\mu$m beams the difference between the position angles was
on average 0.07$^\circ$. In Figure~\ref{fig:rescomp} we show the FWHM
fits to beams on one of the MIGA fields of Table~\ref{tbl:restest}
along with the same data for the corresponding IGA field.  MIGA has a
better resolution than the IGA.  The resolution between the two
MIGA bands is very well matched at all locations in the image. The
situation is quite different for the 60 and 100 $\mu$m bands of the
IGA where the resolution and the position angle of the beams varies
considerably between the two bands.  In a test of 756 simulated 100
and 60 $\mu$m beams the difference between the position angles was on average
7$^\circ$. The reason for this behaviour is that the mid-infrared
detectors on the IRAS focal plane have the same physical size, whereas
the sizes of the far-infrared detectors differ, and that the mid-infrared
detectors are  packed more closely together in the focal plane than
the far-infrared detectors so that their scan pattern on the sky is more
similar (see the IRAS Explanatory Supplement [1988] for details on the
IRAS  detectors and the layout of the focal plane).

The resolution achieved in HIRES processing is a function of the
number of iterations, the coverage pattern, and the strength of the
point source relative to the background value \emph{during
processing}. The latter two factors are what cause the scatter in
resolution seen at a given wavelength in Figure~\ref{fig:rescomp}. The
best resolution, for a given coverage, is achieved for a high ratio of
point source strength to background which is why a bias level is
applied to the image during HIRES processing to bring the background
level of the image as close to zero as possible (see \S~3.5 of Cao et
al.\ [1997] for more details regarding the calculation of the flux
bias).  The bias level that is applied to a given MIGA image is
reported in the image header as a flux in units of W m$^{-2}$. This
value can be converted to an intensity (in Jy sr$^{-1}$) by dividing
the value by the average detector solid angle ($3.2\times10^{-7}$ sr
at 12 $\mu$m and $3.5\times10^{-7}$ sr at 25 $\mu$m [\cite{mos92}])
and by a factor that accounts for the IRAS bandpass shape
($1.348\times10^{-13}$ and $5.16\times10^{-14}$ (Hz) at 12 and 25 $\mu$m 
respectively).

In order to examine the variation of resolution with changing point
source to local processing background ratio (PS/BG) for the MIGA we created
simulated beam maps for the $1.4^{\circ} \times 1.4^{\circ}$ field centered at
$l=74^{\circ}$, $b=-6^{\circ}$ using unscaled point sources with
fluxes of 0.05 to 10000 Jy.  In this region the average background
level was  fairly low,  2.32 MJy sr$^{-1}$ and 6.43 MJy sr$^{-1}$ at
12 $\mu$m and 25 $\mu$m  respectively.  These values can be converted
to a flux using the average detector solid angles yielding average
background fluxes of 0.74 and 2.2 Jy respectively.  To simplify the
test we did not apply a flux bias during processing and so these were
the background values for HIRES processing.  The results of this test
are shown in Figure~\ref{fig:psres}.  These two plots illustrate that
the resolution of a point source is not dependent on just the strength
of the source but the ratio PS/BG.  In the upper plot one sees that for most of
the point sources in this test the resolution achieved at 12 $\mu$m is
slightly better than at 25 $\mu$m, but this is only because of the
lower processing background used at 12 $\mu$m.  In the lower panel we
illustrate this directly by plotting the same resolution measurements
for both 12 and 25 $\mu$m against PS/BG.  All of the data follow the
expected trend of increasing resolution with increasing point source
strength to background ratio.

In Figure~\ref{fig:psbg} we replot the data from
Figure~\ref{fig:rescomp} now showing the FWHM achieved as a function of
PS/BG.  The difference in resolution between the far and
mid-infrared IRAS bands, and the match in resolution between the 12
and 25 $\mu$m bands,  holds over a wide range in
PS/BG. Figure~\ref{fig:psbg} also shows that the trend of increasing
resolution with increasing PS/BG is very flat, so that point
sources with a range of PS/BG will still have similar
achieved resolutions at 12 and 25 $\mu$m.  Although the beam
simulations are done with a realistic assessment of the processing
background for the actual image, still the resolutions reported in the
*fwhm.txt files are for a particular injected point source and so are
only representative of the resolution in the actual images.

\subsection{Ratio Maps and Cross-beam Simulation}
\label{sec:xbs}
This similarity in resolution between the two bands means that, with
care, high resolution ratio maps can be created directly using the
MIGA images.  Care is required because, although the FWHM fits to the
simulated beams are very close, the actual beam shapes are not 2-D
Gaussians and can vary in detail because of the actual PS/BG in the
two images.  As an example, Figure~\ref{fig:ratiomaps} shows two ratio
maps of the region around the HII region S151, chosen because of the
unusual irregular cross-shaped beam pattern caused by the significant
difference in the scan angle between the two IRAS coverages of the
region (this effect is more noticeable at high ecliptic latitudes).  The first
image was constructed by simply dividing the 12 $\mu$m MIGA image by
the 25 $\mu$m MIGA image. 

The other ratio map was constructed using a technique called
cross-band simulation (\cite{fow94}).  This technique makes use
of the HIRES IRAS simulator mode.  First the simulator scans the 12
$\mu$m HIRES image with the 25 $\mu$m detector pattern to create a
simulated view of the 12 $\mu$m sky.  These ``observations'' are then
regularly HIRES processed to create a somewhat lower resolution
version of the 12~$\mu$m image.  The same process is followed for the
25 $\mu$m data. The final result is two images at the same resolution
(slightly poorer than the original 25 $\mu$m) with almost identical
beam shapes. The effect of bringing the two wavebands to the same beam
shape is seen in the way the point sources (cross-shaped beam pattern)
in Figure~\ref{fig:ratiomaps} are undistorted in the cross-beam
simulator image while the point sources show some non-physical
structure in the original.  

Note that the cross-beam simulation does not compensate for differing
PS/BG that may occur between the two bands.  For the most demanding
work, say in examining a particular part of an image with vastly
different PS/BG, flux biases could be chose to better match PS/BG locally.
However, for some purposes even the simple MIGA ratio map may suffice.
The user should closely inspect the simulated beam maps to determine
if the beam shapes are close enough for their purposes.  If so, the common
resolution of the MIGA provides a quick and easy way to obtain high resolution
mid-infrared ratio maps.  Note that any simple ratio map involving the
IGA will not be as satisfactory, and cross-beam simulation will be
desirable.  Therefore we are developing an algorithm to do this using
(M)IGA and its ancillary data, rather than having to return to the raw
IRAS data.

\subsection{Photometry}
\label{sec:photometry}

In order to test the photometric accuracy achieved in the MIGA images
we selected 52 point sources scattered across the range of the atlas.
Sources selected were bright ($>10$ Jy in each band), isolated, and
resolved (as indicated by the point source correlation coefficient in
the IRAS Point Source Catalog (PSC)).  While the Maximum-Correlation
Method (MCM; \cite{aum90}) algorithm at the heart of HIRES conserves
flux globally the flux can be redistributed across the image with each
iteration causing changes in the measured flux at a given location.

Photometry was done using a script driving the IPAC Skyview
program\footnote{Skyview is a general purpose FITS image viewing and
analysis tool available at http://www.ipac.caltech.edu/Skyview/}.
Circular apertures of radius $5'$ and $7'$ were drawn around each
point source and a background surface brightness was defined using the
average of twenty points evenly distributed throughout the annulus
between the two circles.  Flux values were calculated using the two
apertures and compared.  If there was a large discrepancy between the
two values then the background surface brightness was too variable and the
point source was rejected from the sample.  Otherwise the average of
the two fluxes was  used in the comparison with the PSC.  The results
of the photometry are tabulated in Table~\ref{tbl:photometry} and
average values, along with the standard deviation about the average,
are listed in Table~\ref{tbl:avgphoto}.

Both wavelength bands experience the same general trend.  There is an
average 6\% positive offset from the PSC at the $1^{\mathrm st}$
iteration and negative 4\% offset after 20 iterations. This
decrease in point-source flux with iteration is not a universal
property of HIRES processing.  Cao et al. (1997) report for the IGA,
using a sample of 35 point sources, that at  60 $\mu$m, on average
there are 12\% and 14\% offsets from the PSC after 1 and 20 iterations
respectively, and at 100 $\mu$m the offsets are 1\% and 11\%.   In the
IGA case part of the trend is thought to come from a systematic
decrease in the background attributed to increased point-source
ringing, and so one possible cause of the trend observed for the MIGA
might be a systematic increase of the background level with increasing
iterations.  However, as shown in  Table~\ref{tbl:avgphoto} the
average background value actually decreases very slightly. Also since
we do not see this decrease occurring in every single point source
tested (e.g., PSC~20282+3604 or PSC~23239+5754) it is not a universal 
property of the algorithm being used.  

Since this behaviour is different than that seen in the IGA, we
compared a subset of 16 point sources processed using the
ringing reduction algorithm and without (as for the IGA). The results
are shown in Table~\ref{tbl:photoringtest} for the 20$^{\mathrm th}$ 
iteration images. The 1$^{\mathrm st}$ iteration images are
identical in both cases and were remeasured to gain an idea of the
uncertainties involved in the photometric measuring technique being
used. Differences in the measured fluxes were $<0.1\%$; of course a more
sophisticated photometry routine would reduce this
uncertainty further. In general the fluxes measured without the
ringing suppression algorithm in place are higher than the fluxes from
the MIGA images.  

Average results for this test are shown in
Table~\ref{tbl:avgphotoringtest}. Without the ringing suppression
algorithm the point source flux tends to increase as the iterations
increase, like the IGA though less dramatically, whereas with the
algorithm in place the point source flux
tends to decrease with increasing iterations as before. However, this behaviour
is not found for every single source; if the point source flux does
increase for a MIGA source, then it tends to increase less than 
it does when the ringing suppression algorithm is not used, thus
preserving the sense of the relative behaviour. 

While the photometry obtained without the ringing suppression
algorithm in place tends to match the PSC flux values better, we
decided that the possible benefits of having the ringing suppression
algorithm in place outweighed the slightly worse (but still comparing well
to the IGA) photometric performance.  Furthermore, as discussed in
\S~\ref{sec:fluxerr} there are other sources of  error in flux measurement 
that will tend to swamp this uncertainty.

\subsection{Size-Dependent Flux Calibration}
\label{sec:acdc}
One quirk of the IRAS detectors was that their sensitivity was a
function of the dwell-time of a source: the so-called AC/DC effect.
Due to this behaviour two calibrations for IRAS data were developed.
The AC calibration, used for the MIGA and IGA, is suitable for point
sources.  The DC calibration is suitable for measuring fluxes from
extended emission $>2^{\circ}$ in extent (\cite{whe94}).  For structure
at intermediate scales ($6'$ --  $2^{\circ}$) well defined conversion
factors exist for the mid-infrared IRAS bands and users should consult
Table II.B.1 in Wheelock et al.\ (1994).

In order to convert MIGA data to the DC scale, images need to be
multiplied by 0.78 and 0.82 at 12 and 25 $\mu$m, respectively.  Unlike
the 60 and 100 $\mu$m bands, where the correction depends upon the
strength of the source (especially for very bright extended emission),
the AC/DC correction for the mid-infrared IRAS bands appears to be
more stable and does not appear to have any dependence on the source
brightness (IRAS Explanatory Supplement 1988).
 
\subsection{Flux and Surface Brightness Measurement Uncertainty}
\label{sec:fluxerr}

Section 2.3 of Fich \& Terebey (1996) gives a good general discussion
of the variety of factors involved in estimating uncertainties in IRAS
flux measurements, including instrumental, choice of measurement
technique, and background effects, and is recommended reading for users of both
the MIGA and IGA.  It is clear that actual uncertainties in measured point
source flux values from the MIGA will be much higher than the basic 
uncertainty of $\stackrel{<}{_\sim}6$\% implied by the photometric
tests against the PSC discussed in \S~\ref{sec:photometry}.  Due to
complex background emission, uncertainties in flux measurements can
reach as high as 44\% and 20\% at 12 and 25 $\mu$m, respectively. 

Instrumental uncertainties in the mid-infrared IRAS data are smaller than
uncertainties caused by uncertain background estimation and/or
different measurement techniques.  As mentioned in \S~\ref{sec:acdc},
the AC/DC effect, although larger than at far-infrared wavelengths, is
well behaved.  The absolute calibration between IRAS and DIRBE agrees
to 6\% at 12 $\mu$m and 1\% at 25 $\mu$m (\cite{whe94}), illustrating
how the absolute flux calibration for IRAS  is well understood in the 
mid-infrared.

Like for the IGA, the ISSA was used as a large scale surface brightness
truth table for the MIGA. In order to check this
calibration we selected five MIGA images from across the atlas region. The
images were reprojected and rebinned to the geometry and pixel size
of the ISSA and compared pixel-by-pixel with the respective ISSA
plate.  The AC/DC correction was applied to the ISSA plate before the
comparison was made.  The results of the test are tabulated in
Table~\ref{tbl:bright}, and data for one of the test areas are plotted in
Figure~\ref{fig:bright}.

 Any
uncertainties inherent to the ISSA can be passed along to the MIGA.  Of most
concern is the uncertainty in the zodiacal light model that was
subtracted from the ISSA.  Residual zodiacal light removal errors are 
3--5\% of the original emission (0.5 MJy sr$^{-1}$ at 12 $\mu$m and
1.0 MJy sr$^{-1}$ at 25 $\mu$m ) for $\beta > 50^\circ$, and are 1.0 MJy
sr$^{-1}$ at 12 $\mu$m and 2.0--2.5 MJy sr$^{-1}$ over scales of
$10^\circ$ for $50^\circ > \beta > 20^\circ$ (\cite{whe94}). Most of
the MIGA coverage is at high ecliptic latitude and so these residual
errors will be minimal.

\subsection{Positional Accuracy}
\label{sec:positions}

We tested the positional accuracy of the MIGA against the PSC using
the same point sources as for the photometry test.  Using the IPAC
Skyview program a $5'$ radius circle was drawn around the position of
the point source as given in the PSC (which is a weighted average of the 
position at each wavelength), and the flux-weighted centroid
was then measured for the pixels within the circle.  This value was taken
as the position of the point source from the MIGA.

The results of this test are shown in Table~\ref{tbl:pos}. At 12
$\mu$m the average measured distance from the PSC position was 7.88$''
\pm 4.0$ ($\pm 1\sigma$ scatter) and at 25 $\mu$m was 6.69$''\pm3.2$. The
positional accuracy of the MIGA is similar to that reported for the
IGA: 7.6$''\pm5.6$ at 60 $\mu$m and  7.1$''\pm4.1$ at 100 $\mu$m. The
position angle of the offset is different for the 12 and 25 $\mu$m
point sources, but no systematic trend was found.  The position angle
differences are most likely due to a combination of the differences in
the detailed beam shape and the backgrounds in each band which causes the
flux-weighted centroid of the aperture to shift slightly between each band.

To investigate what effect rebinning the MIGA to produce CGPS mosaics
with  $18''$ pixels has on the positional accuracy we repeated the
test on eleven point sources in the W5 region.  The results are shown
in Table~\ref{tbl:pos15} and Table~\ref{tbl:pos18}. Both the original
MIGA and the CGPS mosaiced MIGA are in close agreement, with the
original MIGA agreeing only slightly better with the point source
catalog positions.  At 12 $\mu$m the average distance from the PSC
positions was 6.89$''\pm2.8$ with 15$''$ pixels and 7.32$''\pm4.0$
with 18$''$ pixels.  At 25 $\mu$m the average distance from the PSC
positions was 5.76$''\pm2.3$ and 6.57$''\pm3.5$ with 15$''$ pixels and
18$''$ pixels, respectively. 

\subsection{Mosaic Property}
\label{sec:mosaic}

The large angular scale preprocessing of the IRAS  data allows large
high-quality mosaics to be constructed from the final HIRES images. In
order to quantify the quality of the mosaics, which tend to be
seamless to the eye, we tested the mosaic property using data from four CRDD
plates. This allowed us to study 134 boundaries between images within
the plates and 19 boundaries across plates.  The latter are expected to
be worse because they were preprocessed completely separately.

The tests were done by comparing the pixels along a one pixel
wide strip, one degree long, that is common between adjoining images.
A total of 32294 pixels was examined along boundaries contained
entirely within a single CRDD plate, and 4579 pixels were examined in
the cross-CRDD test.

The pixel ratios ($-1$) and the standard deviations are tabulated in
Tables~\ref{tbl:mosall12} and~\ref{tbl:mosall25} for the 12 and 25 
$\mu$m data, respectively. In Table~\ref{tbl:moscomp} the standard deviations
are shown again along with data from the IGA, showing that the mosaic
quality decreases as one moves toward shorter wavelengths.  This is
caused by a combination of increasing resolution and more complex
backgrounds making the images less smooth as one moves into the
mid-infrared.

\subsection{Residual Hysteresis}
\label{sec:hysteresis}

IRAS detectors experience a hysteresis effect after passing over a
bright source.  While the Galactic plane shadowing effect described by
Cao et al.\ (1997) for the IGA does not affect the 12 and 25 $\mu$m
detectors, hysteresis also can cause bright point sources to have tails
associated with them. Although this can occur at any wavelength, it is
most prominent at 12 and 25 $\mu$m for point sources brighter than
15-20 Jy (\cite{whe94}).  As shown in Figure~\ref{fig:pstail}, a single
source can have a number of tails, one for each scan direction.
Clearly this effect can cause difficulties in the interpretation of
structure near bright point sources, and additional care must be taken even
when doing photometry of bright point sources.

\section{Artifacts}
\label{sec:artifacts}

In the following paragraphs we will discuss briefly various processing
artifacts, namely stripes, glitches, coverage artifacts, and
discontinuities, of which users of the MIGA should be aware.  Since the
processing of the MIGA closely followed that of the IGA, we refer the
reader to Cao et al.\ (1997) for a detailed discussion.  The effect of a new
ringing suppression algorithm is treated in \S~\ref{sec:rings}.

The MIGA used the same destriping technique as in the IGA.  Stripes,
which were once the most common artifact in images constructed from
IRAS data, are now almost completely eliminated from HIRES images.
Details on the destriping algorithm and the application to the IGA can
be found in Cao et al.\ (1996) and Cao et al.\ (1997).

A glitch is the term given to a cosmic-ray or trapped energetic
particle hit on the IRAS detectors that shows up in the IRAS  data
stream.  As with the IGA, the LAUNDR preprocessing
program was used to flag and remove most of the glitches.  It is
possible that some glitches did slip through this stage of the
processing although none have been identified so far. For an
example of a glitch artifact in the IGA see Figure 12 of Cao et al.\ (1997)
(glitches in the MIGA [and EIGA] would have the same properties).

Regions of low IRAS  detector coverage can
cause spurious structure to appear in HIRES images.  Low coverage or
a steep gradient in the coverage can also cause point source positions
to shift.  Users of any HIRES product can use the coverage maps
(cvg\_*, see Table~\ref{tbl:files} and Figure~\ref{fig:allimage_1}) to help
determine if observed features could be affected or even caused by
regions of low coverage.

Discontinuities can occasionally exist entirely within a MIGA image or
mosaic as opposed to across image boundaries.  These discontinuities
trace their origin to the preprocessing step involving the ISSA images (for
calibration and zodiacal emission removal). ISSA data are required corresponding to
the geometry of the input CRDD plate and so mosaics of ISSA
images were constructed if necessary.  Care was taken to minimize the
discontinuities between ISSA images but in some cases a small (on the order of 
1 MJy sr$^{-1}$) 
discontinuity was unavoidable.  This type of discontinuity is usually
not visible in the first iteration image, but is sharpened by the
HIRES algorithm and becomes visible in the twentieth iteration
image. An example of this type of discontinuity is given in Figure 13
of Cao et al.\ (1997).

\subsection{Ringing}
\label{sec:rings}

We were able to apply a ringing suppression algorithm based on Burg
entropy minimization (\cite{cao99}) to the MIGA data.  Ringing is still visible
around point sources but the level of the ringing is significantly
reduced.  Comparison between images constructed with and without the
algorithm showed little difference in their global properties. For
example, Table~\ref{tbl:bnbbright} shows results from a  surface
brightness accuracy test like the tests vs. ISSA presented in
\S~\ref{sec:fluxerr}. In this case we examined five different 10$'$
radius circular apertures at different locations across a 12 $\mu$m
image.  Since these tests showed that the images are virtually
indistinguishable from regularly processed HIRES images away from
point sources, the point source photometry tests in
\S~\ref{sec:photometry} were acceptable (see
Tables~\ref{tbl:photoringtest} and~\ref{tbl:avgphotoringtest}), and
the algorithm had been shown to be useful in at least one published
study (\cite{nor97}), the ringing suppression algorithm was adopted
for MIGA processing.

To quantify the beneficial effect of the ringing suppression algorithm on point
sources in MIGA  we processed five regions without the ringing
suppression algorithm on (WRS) and selected ten well defined, isolated
point sources.  Cuts were then taken across the point sources and the
depth of the ringing on either side of each source was measured
relative to the local background on either side. The change in the
depth of the ringing was then calculated. The effect of the algorithm
is  generally to reduce the depth of the rings in every case. 
Figure~\ref{fig:psrr} illustrates the effect of the algorithm on three
of the point sources at 12 and 25 $\mu$m.  There is also an increase
the peak brightness (not shown). Note the excellent agreement between
the two images as one moves away from the rings. 

We found that as the flux of the point source increases the ringing
tends to become more severe, as might be expected.  At the same
time the absolute value of the change in the ring depth also increases for
stronger point sources. The result is that the fractional reduction in
the ringing does not exhibit any trend with point source flux. The
results of this calculation, along with data on the point source
fluxes, are summarized in Table~\ref{tbl:psrr}. For the point sources 
examined, the average value of ring depth (MIGA/WRS) was 
$0.49\pm0.2$ ($\pm1\sigma$ scatter) and $0.45\pm0.2$ at 12 $\mu$m and 
25 $\mu$m respectively. 

\section {Sample Images}
\label{sec:images}

In this section we display a number of images from MIGA and the CGPS
mosaic version of the MIGA in order to give readers an example of the
data quality.  Colour versions of the images along with other samples
are available on the web\footnote{http::/www.cita.utoronto.ca/$\sim$kerton/MIGA.html}.

In Figure~\ref{fig:issacomp} we contrast 12 and 25 $\mu$m images from ISSA
and the corresponding images from MIGA.  The improvement in the
data quality is obvious.

In Figure~\ref{fig:w5} we show a $4^\circ \times 3^\circ$ mosaic of
the W5 region at 12 and 25 $\mu$m.  This mosaic of twelve MIGA images can
be constructed rapidly since the only operation required on the images
is that they be trimmed before being mosaiced together; no
reprojection step is required. 

On larger scales we show in Figure~\ref{fig:mosaic} one of the CGPS
region mosaics.  The great usefulness of the CGPS data format is that
radio and millimetre data will be available for the same region at the
same geometry and pixel size, greatly facilitating multiwavelength
analysis of objects.  

\section {Summary and Future Directions}
\label{sec:sumfut}

Currently the MIGA covers the CGPS region in longitude and thus
provides a mid-infrared dataset for this multiwavelength survey.  We
considered continuing the MIGA processing to encompass more of the
Galactic plane; however, in the future it is expected that
mid-infrared data from the Mid-Course Space Experiment (MSX;
\cite{pri95}) will be made available for the entire Galactic plane 
($\pm5^\circ$).  Since this data set has a 
higher resolution ($18''$) than is possible to achieve using HIRES, we decided
to focus further expansion of the MIGA to higher and lower Galactic
latitudes in certain key areas tied to expansion of the CGPS.  The CGPS
is expected to enter a second phase of operation starting in 2000,
where the focus will be on disk-halo interaction and extending the
latitude coverage around Cygnus X and the Cepheus star forming
region. In order to study the disk-halo interaction in our
Galaxy effectively obviously one needs to be able to
explore areas above the Galactic plane beyond $\pm6^\circ$.  This
has been clearly demonstrated through investigations of a possible
chimney structure in W4 (\cite{nor96}, \cite{bas99}) and unusual vertical HI
structures (\cite{eng98}) that are analogous to HI ``worms''
(\cite{hei84}). 

The MIGA and IGA are flexible enough that new 
images can be attached seamlessly to existing images due to their
being based on an all-sky survey and the processing technique. In 
Figure~\ref{fig:eiga} we illustrate the addition of the
EIGA images to the original IGA.  This is a nice demonstration of the
ease with which both the IGA and MIGA can be extended to higher or
lower latitudes.
Areas can also be mapped as separate regions using the same
type of processing (e.g., the star-forming regions in Taurus and Ophiucus
that are part of the IGA; IGA also includes Orion which was looked at
by MSX).

Processing of regions outside of the CGPS survey area
is currently underway, starting with the Rho-Oph star forming region
and a HI structure at $l=124^\circ$. We intend to make users aware of
the availability of these data via the CITA web pages\footnote{Latest
information is available at http::/www.cita.utoronto.ca/$\sim$kerton/}. 

As demonstrated in \S~\ref{sec:xbs} cross-band resolution matching is a
useful technique for the construction of large-scale high resolution
color ratio maps.  Unfortunately the means to do this sort of analysis
is not available to the typical user of the MIGA.  At the moment the
only option is to request this sort of processing through IPAC or CITA
(facilities with the IRAS data archive and the requisite software).  We
are currently working on techniques that will allow users to do cross-beam
simulations using the data that comes as part of the MIGA and IGA.
Once available this technique will greatly improve the utility of both
HIRES data sets.

In summary the MIGA provides users of the CGPS data base with a
mid-infrared data set that,  combined with the IGA, will allow users to
study infrared emission from the entire range of dust grain sizes and
thus better study the evolution of dust grains as they move through
different phases of the ISM. Both the MIGA and the IGA can be easily
expanded and built upon to higher and lower Galactic latitudes and
should continue to be useful in the study of disk-halo structures and
star forming regions away from the Galactic plane where higher
resolution infrared data over large angular scales are still not available. 

\acknowledgements

We thank Yu Cao for his assistance and suggestions regarding the
processing  of the MIGA.  Thanks also to Chas Beichman, Ron Beck and
Diane Engler at IPAC for their assistance in obtaining the raw IRAS archive,
and John Fowler for discussions about the HIRES algorithms.
CRK would like to thank the Ontario Graduate Scholarship Program for
support.   This research was supported by the Natural Sciences and
Engineering Research Council of Canada.

\clearpage
\figcaption{Image and beam maps at 12 $\mu$m associated with the
$1.4^\circ \times 1.4^\circ$ area centered at $l=138^\circ$,
$b=2.0^\circ$ (part of the HII region W5 East). The upper row shows
the $1^{\mathrm st}$ (left) and $20^{\mathrm th}$ iteration HIRES
images for the area.  The lower row shows the beam maps for the area.
The background used for processing the beam maps is a smoothed version
of the $20^{\mathrm th}$ iteration HIRES image. \label{fig:allimage_2}}

\figcaption{Ancillary files associated with the same area as in Figure
\ref{fig:allimage_1} . Clockwise from the upper left: det\_*, cvg\_*,
phn\_* (20), and cfv\_* (20).  The detector track map (det\_*) shows
the scan pattern of the IRAS detectors.  The coverage map (cvg\_*) is
created by combining the detector response patterns with the det\_*
map (dark areas show regions of high coverage).  The narrow trim of
low coverage (white) seen in this image is caused by data points that
lie partially outside the processed area and were thus entirely
rejected (images are trimmed before mosaicing to avoid any problems
with such low coverage).  The phn\_* map provides a measure of the
relative noise across an image (black for higher noise).  Its value is
calculated by propagating the a priori noise assigned to a given IRAS
measurement to the image grid.  The cfv\_* map provides a measure of
the relative fitting error, i.e., how much the image has changed from
the previous iteration (black for a larger error/change). A large
value indicates poor, noisy data or saturated regions. A high value
may also indicate that the source is not as fully resolved as possible
(\cite{aum90}; \cite{cao96}) \label{fig:allimage_1}}

\figcaption[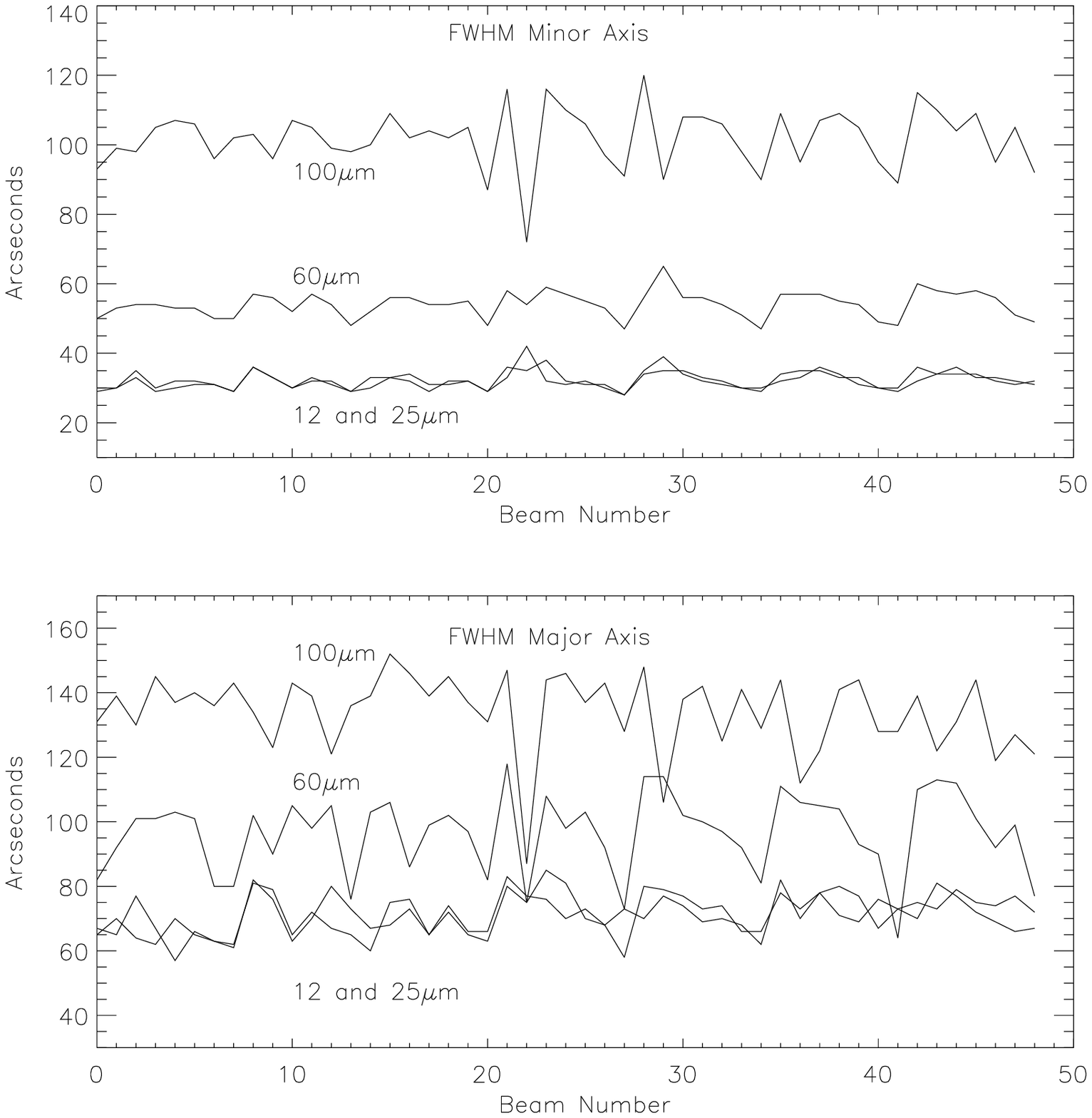]{Comparison of the achieved resolution of
MIGA and IGA images for the $1.4^\circ \times 1.4^\circ$ region
centered at $l=115^\circ$, $b=0^\circ$.  Values plotted are the FWHM of
2D Gaussian fits to the simulated HIRES beams at 49 locations
(distributed evenly in a 7$\times$7 grid) in the image.  Notice the
large difference in resolution between the two IGA bands (100 and 60
$\mu$m).  In contrast the resolution of the two MIGA bands (12 and 25
$\mu$m) is closely matched at all locations in the image.  Note that
the point sources used to construct the simulated beams have been
scaled and have been processed at different background levels.  This factor
causes most of the variation in FWHM between 12 and 25 $\mu$m for a
given point source (see text and Fig.~\ref{fig:psbg} for details). 
\label{fig:rescomp}}

\figcaption[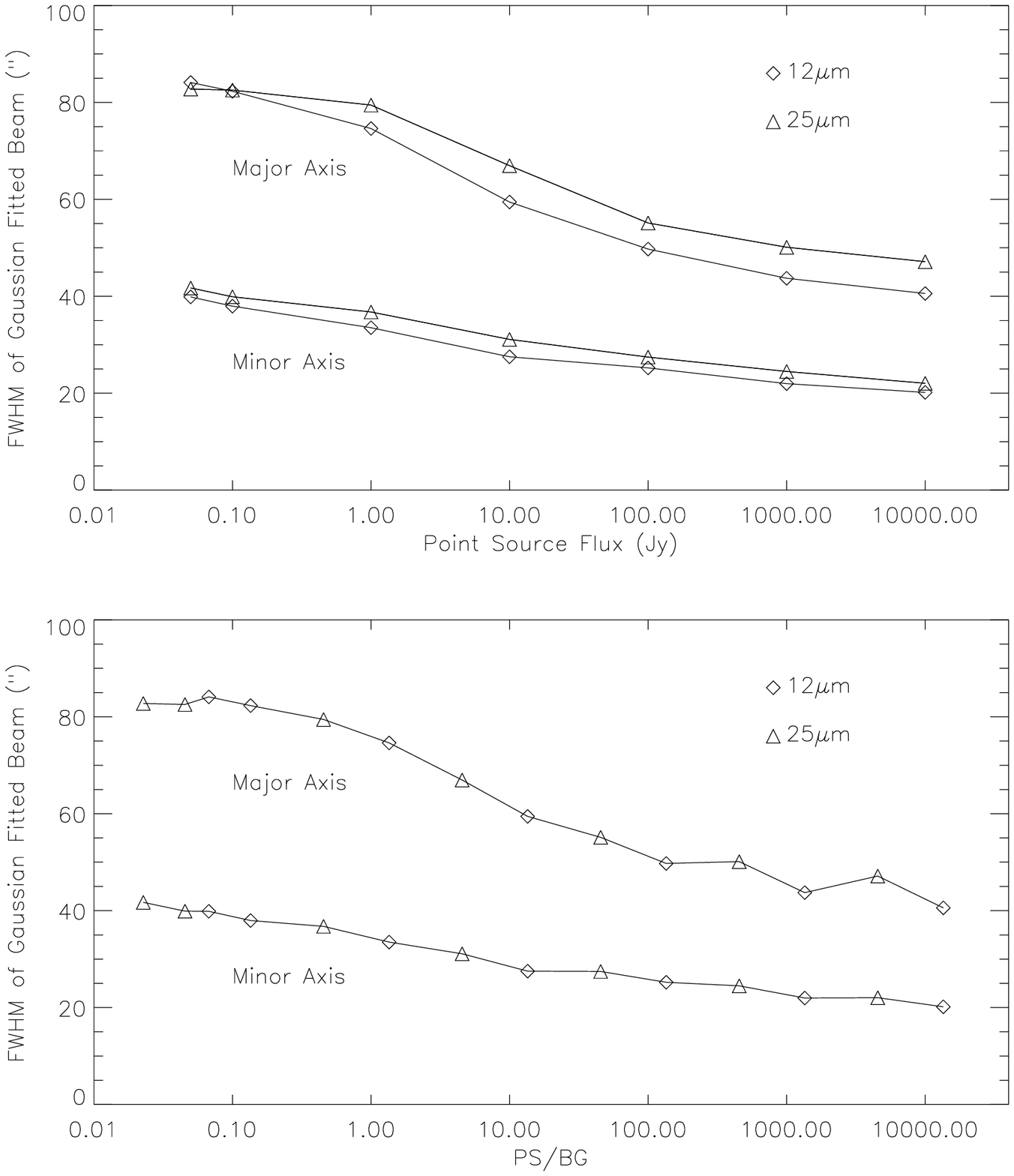]{Dependence of beam resolution on source and
background flux (BG). In the top figure we show the resolution
as a function of the point source strength (PS). At any given PS, the
ratio PS/BG is higher at 12 $\mu$m than at 25 $\mu$m due to the lower
12 $\mu$m background level, and this results in a better resoution at
12 $\mu$m.   In the lower figure we show this directly by plotting the
same data points as a function of the PS/BG.  The resolution of the
HIRES beam improves steadily in both bands as PS/BG increases,
tracking the same locus. Each point shown is the average value for 49 
simulated beams evenly arrayed in a test field located at $l=74^{\circ}$,
$b=-6^{\circ}$. \label{fig:psres}}

\figcaption[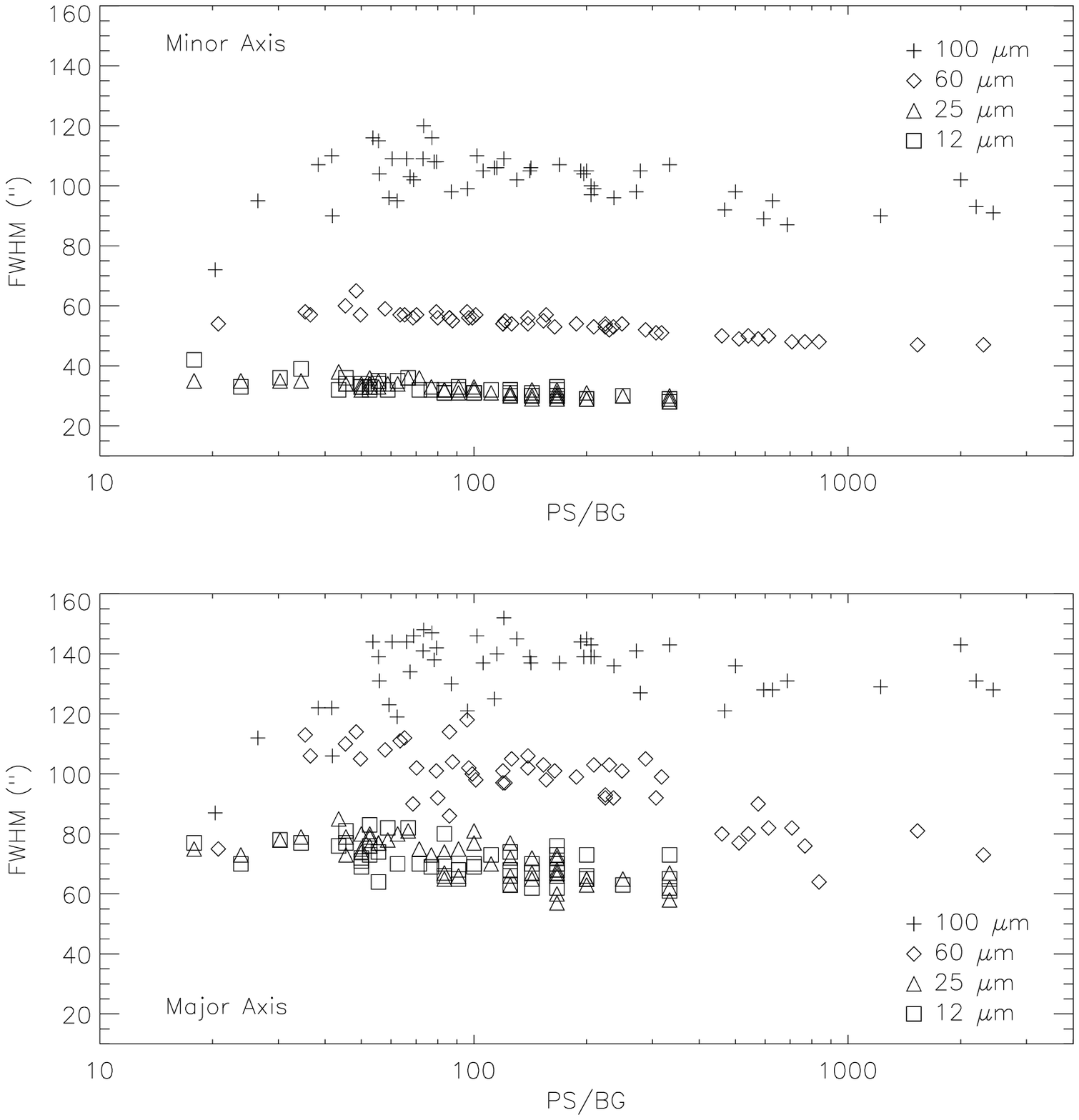]{Data from Figure~\ref{fig:rescomp}
replotted as a function of the ratio of point source flux to
background flux \emph{during HIRES processing} (PS/BG).  The match in
resolution at 12 and 25 $\mu$m holds over a wide range of PS/BG.
\label{fig:psbg}}

\figcaption{Comparison of ratio maps (12/25) constructed using a
simple division of MIGA images (top) and using the HIRES IRAS
Simulator to perform cross-band beam matching (bottom).  Images are
linearly stretched from 0 (white) to 1 (black).  This causes most of
the point sources (stars) to saturate and appear as
black structures since they are brighter at 12 $\mu$m than at 25 $\mu$m.
The two noticeable exceptions are the point source associated with
S151 at $l=108.5$, $b=-2.8$ and the point source at $l=107.6$, $b=-2.24$
(the planetary nebula PK 107$-$2.1). Note that in general structure
shown in each image is the same, but that the point sources are more
clearly defined in the bottom image because of the matching beam
shapes. For example, the feature associated with S151 is more clearly
shown to be a point source in the beam-matched image than in the
simple ratio image. The lower image also has a less mottled background
due to the matched beams and the slightly lower resolution.
\label{fig:ratiomaps}}

\figcaption[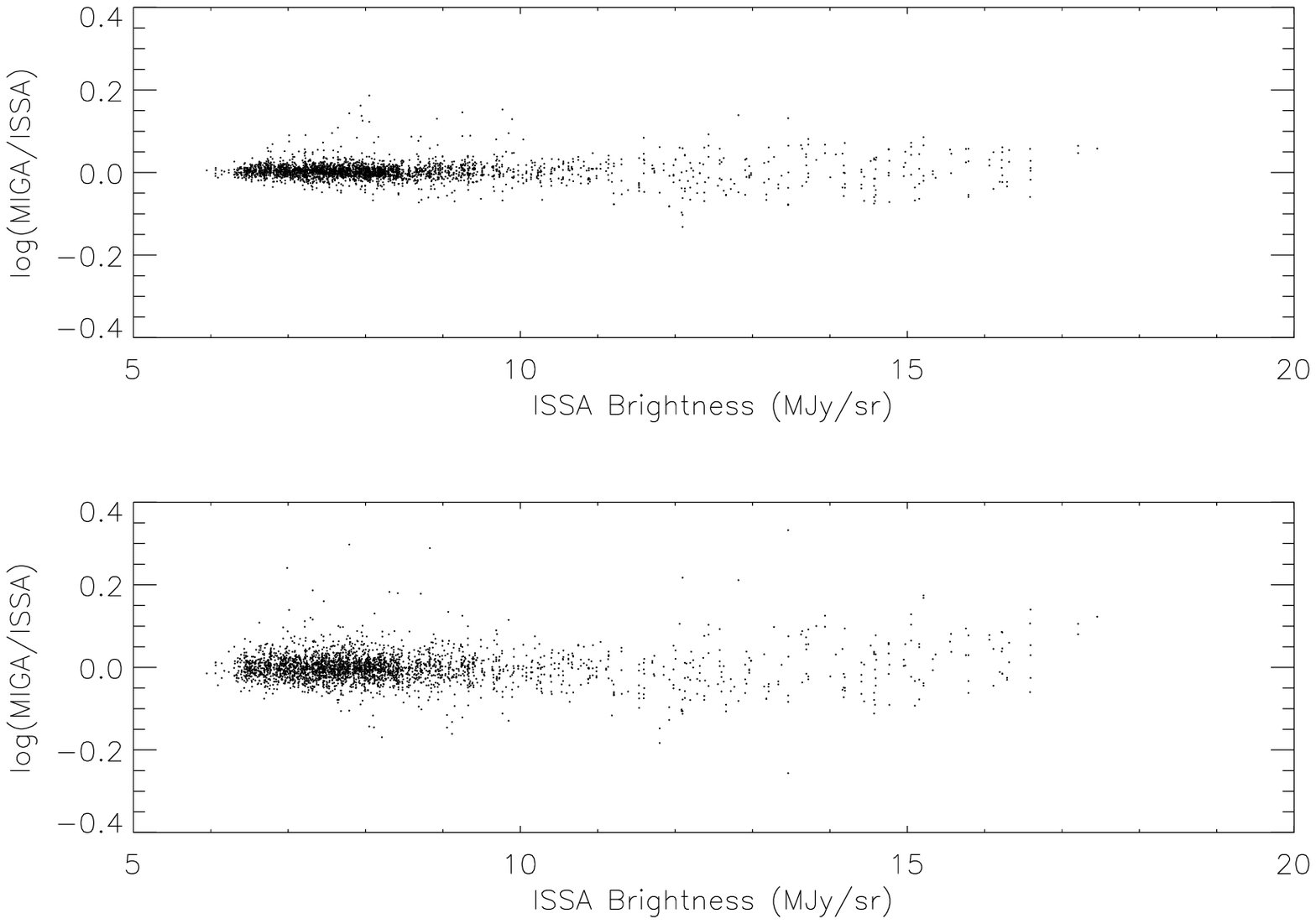]{Data from the surface brightness test for
the 12 $\mu$m MIGA image at $l = 75^\circ$, $b = 2.0^\circ$. Top panel
shows iteration 1, bottom panel shows iteration 20. \label{fig:bright}}

\figcaption{A spectacular demonstration of multiple point source tails
from detector hysteresis.  This 12 $\mu$m MIGA mosaic shows two tails 
extending from the extremely bright W3 region in the lower left corner
of the image.  The image is linearly stretched between 4.44 MJy
sr$^{-1}$(white) and 15.0 MJy sr$^{-1}$ (black).  The offset between
the brightness of the tails and the area outside of them is $\sim$1
MJy sr$^{-1}$ over most of the length of the tail. Two tails are
present because of the two noticeably different scan directions of the
IRAS satellite in this region. \label{fig:pstail}}

\figcaption[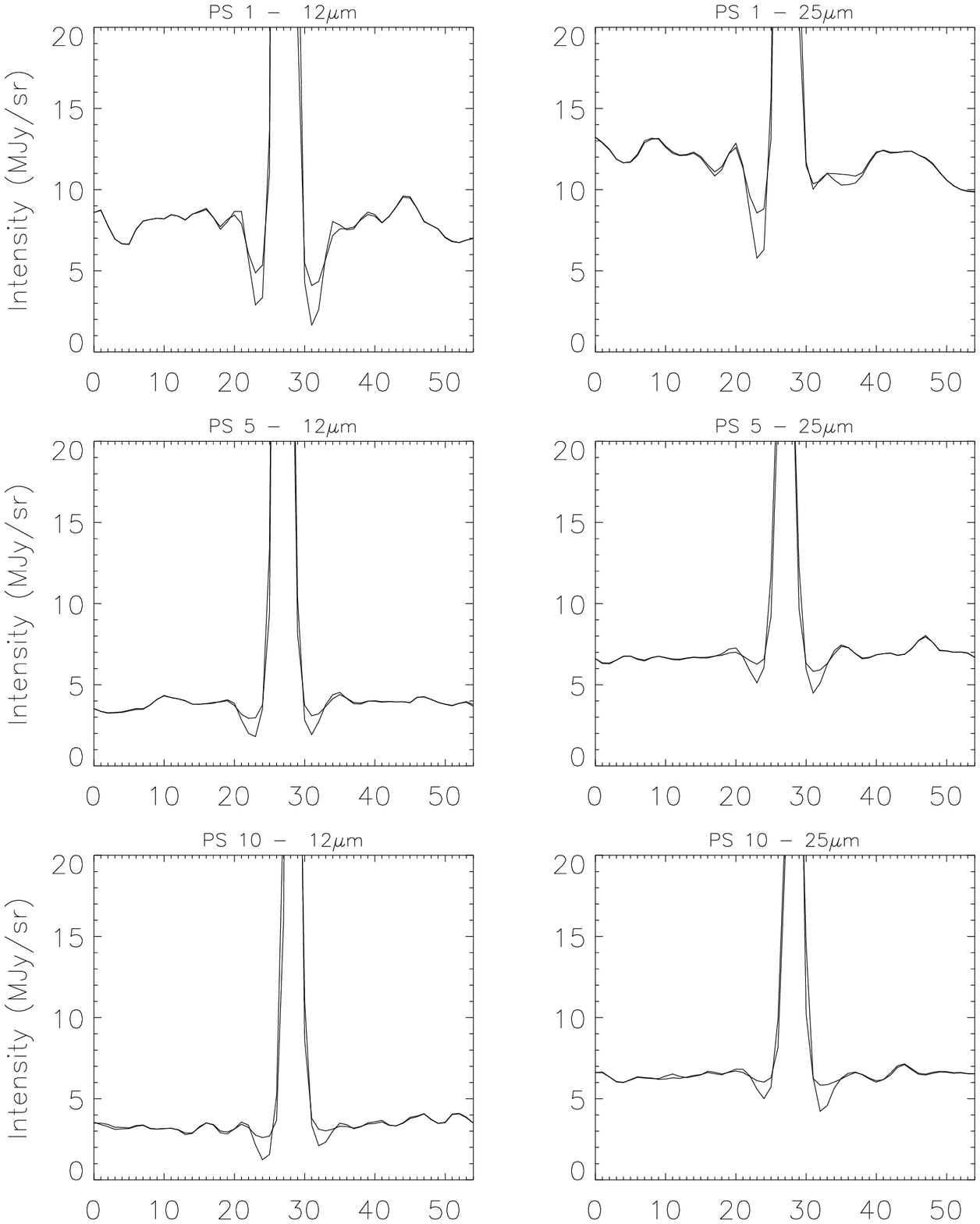]{Illustration of the point source ringing
suppression algorithm for three of the observed point sources at 12
and 25 $\mu$m (left and right panels, respectively).  The figures show
cuts taken across the center of the point source after
processing with and without the algorithm.  For simplicity, cuts were
taken at constant Galactic latitude and thus the angle between the
cut, and the minor axis of the point source and the scan direction(s) is
variable. In all cases the algorithm led to a reduction in the amount
of ringing around the point source, and so the deeper troughs are
associated with processing without the algorithm. The peak 12 $\mu$m
intensities, without the ringing suppression algorithm, are 308, 105 and
59 MJy sr$^{-1}$ for point sources 1, 5 and 10, respectively.  At 25
$\mu$m, without ringing suppression, the peak intensities are 86, 41
and 37 MJy sr$^{-1}$, respectively. The amount of ringing depends on
the peak intensity (see text for details). \label{fig:psrr}}

\figcaption{Comparison of ISSA and MIGA images showing mid-infrared emission
associated with the HII region W5-East (12 $\mu$m at top and
25 $\mu$m at bottom). The MIGA images, on the
left, were cropped from MIGA mosaics made for Fig.~\ref{fig:w5}.  At
12 $\mu$m most of the emission is associated with the edges of
molecular clouds surrounding the central ionized region. At 25 $\mu$m
emission also extends into the ionized region. Both sets of images have the
same logarithmic stretch from 3.0 -- 700 MJy sr$^{-1}$ (white -- black). 
\label{fig:issacomp}}

\figcaption{Two $4^\circ \times 3^\circ$ mosaics of the W5 star
forming region (12 $\mu$m at top and
25 $\mu$m at bottom).  The top image is linearly stretched from 2.7 --
16.1 MJy sr$^{-1}$ (white -- black) and the bottom image is linearly
stretched from 6.0 -- 29.2 MJy sr$^{-1}$.  \label{fig:w5}}

\figcaption{CGPS $5.12^{\circ} \times 5.12^{\circ}$ mosaics
demonstrating the utility of the MIGA in studying ISM structures of
large angular extent (12 $\mu$m at top and 25 $\mu$m at bottom).  The
feature in the center of the mosaic is infrared emission associated
with the HII region LBN 140.8$-$1.4 (BFS 27 in \cite{bfs82}). The top
image is linearly stretched from 1.8 -- 15 MJy sr$^{-1}$, and the
bottom image is stretched from 5.3 -- 17 MJy sr$^{-1}$ (white --
black).  \label{fig:mosaic}}

\figcaption{The Extended IGA (EIGA).  The top image shows a 100 $\mu$m
CGPS mosaic region with the original extent of the IGA data.  In order
to provide full far-infrared coverage for the CGPS we constructed an
extension to the IGA.  The bottom image shows the same mosaic with the
EIGA added on.  Agreement between the IGA and EIGA was excellent.  The
applied linear stretch is from 24 -- 120 MJy sr$^{-1}$ (top) and 22 --
112 MJy sr$^{-1}$ (bottom) (white -- black).  \label{fig:eiga}}

\clearpage

\begin{deluxetable}{ll}
\footnotesize
\tablecaption{MIGA Atlas Files \label{tbl:files}}
\tablewidth{0pt}
\tablehead{
\colhead{File Type\tablenotemark{1}}  &  \colhead{Description}
}
\startdata
bem\_*		& Image showing location of point sources used to generate img\_*bem* files \nl
cfv\_*		& Correction Factor Variance, $1^{\mathrm st}$ and $20^{\mathrm th}$ iteration \nl
cvg\_*		& Coverage \nl
det\_*		& Detector Tracks \nl
img\_*		& $1^{\mathrm st}$ and $20^{\mathrm th}$ iteration
HIRES images \nl
img\_*bem*	& Beam maps, $1^{\mathrm st}$ and $20^{\mathrm th}$ iteration \nl
phn\_*		& Photometric Noise, $1^{\mathrm st}$ and $20^{\mathrm th}$ iteration \nl
*fwhm.txt	& Tabulation of 2-D Gaussian fits to $20^{\mathrm th}$ iteration beams \nl
\enddata
\tablenotetext{1}{The * refers to text in the file name that gives the image
wavelength (12 or 25 $\mu$m), iteration, and/or location (CRRD plate number and
Galactic coordinates)}
\end{deluxetable}
\clearpage

\begin{deluxetable}{lcccccccc}
\footnotesize
\tablecaption{Resolution Test \label{tbl:restest}}
\tablewidth{0pt}
\tablehead{ 
\colhead{} & \multicolumn{2}{c}{Major Axis - 12 $\mu$m} &  \multicolumn{2}{c}{Minor Axis - 12 $\mu$m} & \multicolumn{2}{c}{Major Axis - 25 $\mu$m} &  \multicolumn{2}{c}{Minor Axis - 25 $\mu$m} \nl 
\colhead{Field\tablenotemark{1}} & \colhead{FWHM\tablenotemark{2} ($''$)} & \colhead{St. Dev.} & \colhead{FWHM ($''$)} & \colhead{St. Dev.} & \colhead{FWHM ($''$)} & \colhead{St. Dev.} & \colhead{FWHM ($''$)} & \colhead{St. Dev.} 
}
\startdata
g075$+$02 & 74.24 & \phantom{1}5.48 & 33.65 & 2.70 & 72.81 & \phantom{1}6.25 & 34.24 & 2.44 \nl
g077$-$04 & 76.37 & \phantom{1}4.06 & 34.98 & 2.17 & 75.29 & \phantom{1}3.76 & 35.86 & 1.74 \nl
g084$+$03 & 51.55 & 10.4 & 41.33 & 5.51 & 50.02 & 10.3 & 39.31 & 4.82 \nl
g094$-$06 & 57.59 & \phantom{1}5.52 & 28.14 & 2.38 & 58.76 & \phantom{1}7.48 & 29.69 & 3.30 \nl
g107$+$05 & 55.45 & \phantom{1}6.79 & 33.33 & 2.84 & 57.12 & \phantom{1}7.27 & 34.82 & 3.15 \nl
g133$+$00 & 70.90 & \phantom{1}5.61 & 32.14 & 2.65 & 71.67 & \phantom{1}6.77 & 32.18 & 2.31 \nl
g115$-$01 & 67.63 & \phantom{1}6.13 & 32.41 & 2.10 & 66.80 & \phantom{1}6.37 & 33.02 & 2.01 \nl
g116$-$02 & 64.73 & \phantom{1}6.74 & 31.43 & 2.08 & 64.29 & \phantom{1}6.80 & 32.39 & 2.46 \nl
g118$+$02 & 66.57 & \phantom{1}4.78 & 32.27 & 1.99 & 64.65 & \phantom{1}6.04 & 32.67 & 2.94 \nl
g119$-$01 & 69.55 & \phantom{1}5.59 & 32.90 & 1.77 & 70.22 & \phantom{1}5.74 & 33.59 & 2.23 \nl
g123$-$06 & 64.96 & \phantom{1}6.40 & 29.98 & 2.20 & 60.47 & \phantom{1}8.15 & 29.90 & 2.02 \nl
g138$+$04 & 80.57 & \phantom{1}4.28 & 35.73 & 2.08 & 82.51 & \phantom{1}3.64 & 36.65 & 2.15 \nl
\nl
Average\tablenotemark{3} & 66.7$\pm$2.5  &  & 33.2$\pm$0.9  &  & 66.2$\pm$2.6  &  &  33.7$\pm$0.8 \nl
\enddata
\tablenotetext{1}{Field notation indicates the field center in Galactic coordinates}
\tablenotetext{2}{FWHM reported is the average of 49 beams in each
field; their standard deviation is given in the adjacent column}
\tablenotetext{3}{Average$\pm$standard deviation of the mean}

\end{deluxetable}

\clearpage

\begin{deluxetable}{lllllllll}
\footnotesize
\tablecaption{MIGA Photometry Test \label{tbl:photometry}}
\tablewidth{0pt}
\tablehead{
\colhead{} & \multicolumn{2}{c}{PSC Position} & \multicolumn{2}{c}{PSC Flux (Jy)} & 
\multicolumn{4}{c}{MIGA Flux (Jy)} \nl \cline{6-9}
\colhead{Source} & \colhead{$l$} & \colhead{$b$}  & \colhead{12 $\mu$m}
& \colhead{25 $\mu$m}  &\colhead{12 $\mu$m (1)} & \colhead{12 $\mu$m (20)} & \colhead{25 $\mu$m (1)} & \colhead{25 $\mu$m (20)}
}
\startdata
00067+6340 & 118.3409 & \phantom{$-$}1.4576 & 53.36 & \phantom{2}28.22 & 63.5363 & 48.6534 & \phantom{2}28.5863 & \phantom{1}22.7596 \nl
00186+5940 & 119.1762 & $-$2.6994 & 24.51 & \phantom{2}10.14 & 25.7810 & 25.0849 & \phantom{2}10.0811 & \phantom{1}10.9718 \nl
00340+6251 & 121.3003 & \phantom{$-$}0.3074 & 50.41 & \phantom{2}19.37 & 47.0268 & 42.3261 & \phantom{2}19.1033 & \phantom{1}17.0207 \nl
00362+5924 & 121.3715 & $-$3.1555 & 25.50 & \phantom{2}15.20 & 24.6547 & 22.9576 & \phantom{2}15.9971 & \phantom{1}13.6838 \nl
01071+6551 & 124.8524 & \phantom{$-$}3.3233 & 24.35 & \phantom{2}12.30 & 26.0445 & 23.2551 & \phantom{2}12.6349 & \phantom{1}12.5230 \nl
01142+6306 & 125.8542 & \phantom{$-$}0.6440 & 30.06 & \phantom{2}18.08 & 30.4429 & 29.8279 & \phantom{2}18.1147 & \phantom{1}17.0529 \nl
01145+5902 & 126.2820 & $-$3.3949 & 21.64 & \phantom{2}10.30 & 25.9571 & 22.3019 & \phantom{2}12.0768 & \phantom{1}10.3597 \nl
01217+6049 & 126.9839 & $-$1.5298 & 52.31 & \phantom{2}31.26 & 52.7115 & 45.3018 & \phantom{2}32.8308 & \phantom{1}29.0992 \nl
01261+6446 & 126.9464 & \phantom{$-$}2.4612 & 21.84 & \phantom{2}12.04 & 22.0756 & 21.7847 & \phantom{2}13.1082 & \phantom{1}11.9172 \nl
01364+6038 & 128.7872 & $-$1.4185 & 17.41 & \phantom{2}10.25 & 17.0657 & 15.0812 & \phantom{22}9.8891 & \phantom{1}10.9009 \nl
01443+6417 & 128.9559 & \phantom{$-$}2.3395 & 49.95 & \phantom{2}18.87 & 50.0009 & 45.8308 & \phantom{2}19.7812 & \phantom{1}17.6047 \nl
01572+5844 & 131.7681 & $-$2.6935 & 25.62 & \phantom{2}16.11 & 19.8032 & 21.7792 & \phantom{2}18.0632 & \phantom{1}15.6846 \nl
01577+6354 & 130.4745 & \phantom{$-$}2.3018 & 26.53 & \phantom{2}19.41 & 33.4020 & 29.0409 & \phantom{2}19.8670 & \phantom{1}18.1097 \nl
02044+6031 & 132.1572 & $-$0.7257 & 12.05 & 105.8 & 14.2885 & 13.1192 & 115.181 & 106.661 \nl
02086+6355 & 131.6201 & \phantom{$-$}2.6685 & 29.37 & \phantom{2}17.32 & 30.0834 & 26.4275 & \phantom{2}16.5831 & \phantom{1}16.2115 \nl
02217+5712 & 135.3331 & $-$3.1573 & 18.53 & \phantom{2}13.25 & 21.5100 & 17.9358 & \phantom{2}14.5014 & \phantom{1}13.1015 \nl
02347+5649 & 137.1203 & $-$2.8495 & 38.85 & \phantom{2}26.12 & 37.7633 & 38.1541 & \phantom{2}26.5939 & \phantom{1}24.3429 \nl
02469+5646 & 138.6543 & $-$2.2057 & 90.58 & \phantom{2}78.86 & 93.6932 & 77.1910 & \phantom{2}85.3338 & \phantom{1}71.0313 \nl
02473+5738 & 138.3201 & $-$1.3976 & 39.04 & \phantom{2}26.45 & 47.1570 & 38.3443 & \phantom{2}33.5275 & \phantom{1}28.1821 \nl
03042+5850 & 139.7360 & \phantom{$-$}0.6994 & 22.62 & \phantom{2}13.02 & 22.0879 & 18.8522 & \phantom{2}13.9422 & \phantom{1}12.3470 \nl
03094+5530 & 142.0154 & $-$1.8251 & 57.78 & \phantom{2}21.07 & 66.1875 & 55.8660 &\phantom{2}23.7564 & \phantom{1}20.3264 \nl
03301+5658 & 143.6405 & \phantom{$-$}0.9657 & 15.12 & \phantom{2}14.72 & 18.8210 & 17.0154 & \phantom{2}16.2024 & \phantom{1}15.7216 \nl
03385+5927 & 143.0775 & \phantom{$-$}3.6207 & 59.41 & \phantom{2}36.58 & 50.4041 & 42.7497 & \phantom{2}29.2350 & \phantom{1}26.7332 \nl
03419+5429 & 146.4398 & $-$0.0588 & 15.53 & \phantom{2}12.64 & 14.7206 & 12.0248 & \phantom{2}12.6153 & \phantom{1}12.2009 \nl
20026+4018 & \phantom{1}76.4321 & \phantom{$-$}4.8081 & 23.98 & \phantom{2}11.34 & 23.7829 & 22.1433 & \phantom{2}12.4370 & \phantom{1}11.7800 \nl
20028+3910 & \phantom{1}75.4887 & \phantom{$-$}4.1759 & 41.78 & 210.8 & 45.3307 & 42.7668 & 224.879 & 185.296 \nl
20282+3604 & \phantom{1}75.7746 & $-$1.7196 & 39.74 & \phantom{2}17.27 & 39.8012 & 40.5354 & \phantom{2}18.6999 & \phantom{1}19.1740 \nl
20422+4644 & \phantom{1}85.8464 & \phantom{$-$}2.6680 & \phantom{1}14.34 & 16.03 & 14.3901 & 13.4234 & 16.4262 & 14.0029 \nl
20499+4657 & \phantom{1}86.8478 & \phantom{$-$}1.7934 & \phantom{1}29.22 & 25.37 & 29.9273 & 24.1347 & 25.0772 & 21.4368 \nl
20502+4709 & \phantom{1}87.0338 & \phantom{$-$}1.8859 & 104.9 & 62.22 & 110.631 & 91.4837 & 74.4617 & 59.6315 \nl
20549+5245 & \phantom{1}91.7978 & \phantom{$-$}4.9220 & \phantom{1}64.88 & 51.15 & 65.9244 & 61.7231 & 55.3033 & 50.2928 \nl
21015+4859 & \phantom{1}89.6478 & \phantom{$-$}1.6478 & \phantom{1}26.04 & 16.14 & 30.6040 & 29.8599 & 17.3360 & 18.0732 \nl
21122+4900 & \phantom{1}90.8629 & \phantom{$-$}0.3731 & \phantom{1}14.43 & 16.66 & 18.7282 & 17.1637 & 20.5298 & 18.8723 \nl
21167+5502 & \phantom{1}95.6773 & \phantom{$-$}4.0997 & \phantom{1}24.22 & 10.18 & 29.2942 & 27.4664 & 11.2405 & 11.0972 \nl
21223+5114 & \phantom{1}93.5840 & \phantom{$-$}0.8016 & \phantom{1}69.98 & 48.78 & 66.8236 & 66.2346 & 57.7606 & 50.2060 \nl
21232+5705 & \phantom{1}97.7603 & \phantom{$-$}4.9227 & \phantom{1}44.59 & 17.65 & 47.1036 & 41.8459 & 18.2655 & 15.2005 \nl
21282+5050 & \phantom{1}93.9862 & $-$0.1185 & \phantom{1}50.99 & 74.36 & 57.2724 & 41.6584 & 77.5648 & 62.0380 \nl
21444+5053 & \phantom{1}95.9339 & $-$1.7734 & \phantom{1}17.16 & 15.68 & 18.0558 & 17.0845 & 16.2218 & 15.1624 \nl
21453+5959 & 101.8655 & \phantom{$-$}5.1417 & \phantom{1}24.38 & 19.08 & 20.4740 & 17.4520 & 18.1115 & 15.7566 \nl
21475+5211 & \phantom{1}97.1263 & $-$1.0757 & \phantom{1}22.89 & 13.41 & 25.4282 & 24.9830 & 15.4437 & 15.8910 \nl
22122+5745 & 103.3018 & \phantom{$-$}1.2718 & \phantom{1}32.19 & 17.73 & 37.8056 & 32.9208 & 19.7780 & 19.6665 \nl
22248+6058 & 106.3922 & \phantom{$-$}3.0933 &\phantom{1} 12.04 & 12.19 & 12.3764 & 11.6632 & 13.7652 & 13.2977 \nl
22303+5950 & 106.3932 & \phantom{$-$}1.7784 & \phantom{1}54.46 & 68.41 & 61.3254 & 48.0261 & 71.6826 & 62.2615 \nl
22413+5929 & 107.4251 & \phantom{$-$}0.7972 & \phantom{1}58.52 & 28.41 & 55.7869 & 59.0043 & 28.2041 & 27.6202 \nl
22471+5902 & 107.8860 & \phantom{$-$}0.0507 & \phantom{1}19.04 & 14.47 & 18.2206 & 16.9294 & 14.3115 & 14.5127 \nl
22489+6359 & 110.2892 & \phantom{$-$}4.3798 & \phantom{1}46.38 & 22.28 & 46.4414 & 37.4446 & 21.7870 & 17.7950 \nl
23000+5932 & 109.5835 & $-$0.1895 & \phantom{1}55.61 & 38.48 & 57.0682 & 48.7887 & 39.9623 & 32.8122 \nl
23106+6340 & 112.3522 & \phantom{$-$}3.1283 & \phantom{1}30.74 & 14.38 & 38.2095 & 34.1633 & 14.8624 & 13.9579 \nl
23239+5754 & 111.8786 & $-$2.8500 & \phantom{1}20.86 & 71.22 & 23.8860 & 24.2619 & 72.3286 & 72.4818 \nl
23281+5742 & 112.3421 & $-$3.2212 & \phantom{1}83.73 & 69.30 & 89.4587 & 77.0907 & 75.9988 & 61.9112 \nl
23321+6545 & 115.2069 & \phantom{$-$}4.3214 & \phantom{1}13.66 & 85.61 & 15.8241 & 13.2595 & 102.823 & 78.5526 \nl
23592+6228 & 117.2816 & \phantom{$-$}0.4239 & \phantom{1}23.18 & 16.04 & 26.4240 & 22.1920 & 16.8456 & 14.7434 \nl
\enddata
\tablenotetext{}{NOTE---Galactic latitude and longitude ($l,b$) are given in degrees.}
\end{deluxetable}

\clearpage

\begin{deluxetable}{lll}
\tablecaption{MIGA Photometry Test Average Results \label{tbl:avgphoto}}
\tablewidth{0pt}
\tablehead{
\colhead{Ratio} & \colhead{Mean} & \colhead{St. Dev.}
}
\startdata
12 $\mu$m Band: &  & \nl
\hspace*{0.5cm} MIGA(1)/PSC  &  1.06\phantom{0}	& 0.11 \nl
\hspace*{0.5cm} MIGA(20)/PSC &  0.95		& 0.11 \nl
\hspace*{0.5cm} MIGA(20)/MIGA(1) & 0.90	& 0.076 \nl
\hspace*{0.5cm} Mean Background\tablenotemark{\dagger} (1) & 3.94	& 1.57  \nl
\hspace*{0.5cm} Mean Background (20) & 3.89	& 1.57  \nl
25 $\mu$m Band: &  & \nl
\hspace*{0.5cm} MIGA(1)/PSC  &  1.06		& 0.080 \nl
\hspace*{0.5cm} MIGA(20)/PSC &  0.96		& 0.096 \nl
\hspace*{0.5cm} MIGA(20)/MIGA(1) & 0.91	& 0.078 \nl
\hspace*{0.5cm} Mean Background (1) & 6.94	& 1.72  \nl
\hspace*{0.5cm} Mean Background (20) & 6.87	& 1.70  \nl
\enddata
\tablenotetext{\dagger}{Background units are MJy sr$^{-1}$}
\end{deluxetable}

\clearpage

\begin{deluxetable}{lclll}
\tablecaption{Ringing Suppression Photometry Test at Iteration 20 \label{tbl:photoringtest}}
\tablewidth{0pt}
\tablehead{& \multicolumn{2}{c}{12 $\mu$m Flux (Jy)} &
\multicolumn{2}{c}{25 $\mu$m Flux (Jy)} \nl
& \colhead{MIGA}  & \colhead{WRS}  & \colhead{MIGA}
& \colhead{WRS} \nl
\colhead{Source} 
}
\startdata
00186+5940 & 25.0849 & \phantom{1}26.4385 &  \phantom{1}10.9718   & \phantom{2}11.1186  \nl
00340+6251 & 42.3261 & \phantom{1}50.7771 &  \phantom{1}17.0207   & \phantom{2}19.8332  \nl
01142+6306 & 29.8279 & \phantom{1}36.0347 &  \phantom{1}17.0529   & \phantom{2}19.5175  \nl
01572+5844 & 21.7792 & \phantom{1}24.7381 &  \phantom{1}15.6846   & \phantom{2}19.7868  \nl
01577+6354 & 29.0409 & \phantom{1}32.4511 &  \phantom{1}18.1097   & \phantom{2}18.8215  \nl
02086+6355 & 26.4275 & \phantom{1}32.3267 &  \phantom{1}16.2115   & \phantom{2}17.8623  \nl
02469+5646 & 77.1910 & \phantom{1}92.3777 &  \phantom{1}71.0313   & \phantom{2}84.1628  \nl
03385+5927 & 42.7497 & \phantom{1}48.4338 &  \phantom{1}26.7332   & \phantom{2}32.3640  \nl
20028+3910 & 42.7668 & \phantom{1}45.6190 &            185.296    & 220.081 \nl
20499+4657 & 24.1347 & \phantom{1}27.4869 &  \phantom{1}21.4368   & \phantom{2}24.2367  \nl
20502+4709 & 91.4837 &           115.012  &  \phantom{1}59.6315   & \phantom{2}75.2420  \nl
20549+5245 & 61.7231 & \phantom{1}67.7785 &  \phantom{1}50.2928   & \phantom{2}56.2955  \nl
21444+5053 & 17.0845 & \phantom{1}18.5642 &  \phantom{1}15.1624   & \phantom{2}15.8333  \nl
22413+5929 & 59.0043 & \phantom{1}60.1043 &  \phantom{1}27.6202   & \phantom{2}29.4132  \nl
23000+5932 & 48.7887 & \phantom{1}60.8649 &  \phantom{1}32.8122   & \phantom{2}41.6368  \nl
23281+5742 & 77.0907 & \phantom{1}93.1810 &  \phantom{1}61.9112   & \phantom{2}80.3636  \nl

\enddata
\tablenotetext{}{WRS -- Without Ringing Suppression}
\end{deluxetable}

\clearpage

\begin{deluxetable}{lll}
\tablecaption{Ringing Suppression Photometry Test --- Average Results \label{tbl:avgphotoringtest}}
\tablewidth{0pt}
\tablehead{
\colhead{Ratio} & \colhead{Mean} & \colhead{St. Dev.}
}
\startdata
12 $\mu$m Band: &  & \nl
\hspace*{0.5cm} MIGA(1)/PSC      &  1.01   &  0.106  \nl
\hspace*{0.5cm} MIGA(20)/PSC     &  0.92  &  0.097  \nl
\hspace*{0.5cm} MIGA(20)/MIGA(1) &  0.91  &  0.084  \nl
\hspace*{0.5cm} WRS(1)/PSC       &  1.01   &  0.105  \nl
\hspace*{0.5cm} WRS(20)/PSC      &  1.06   &  0.098  \nl
\hspace*{0.5cm} WRS(20)/WRS(1)   &  1.05   &  0.082  \nl
25 $\mu$m Band: &  & \nl
\hspace*{0.5cm} MIGA(1)/PSC      &  1.03  &  0.087  \nl 
\hspace*{0.5cm} MIGA(20)/PSC     &  0.92 &  0.078  \nl
\hspace*{0.5cm} MIGA(20)/MIGA(1) &  0.90 &  0.077  \nl
\hspace*{0.5cm} WRS(1)/PSC       &  1.03  &  0.087  \nl
\hspace*{0.5cm} WRS(20)/PSC      &  1.06  &  0.089  \nl
\hspace*{0.5cm} WRS(20)/WRS(1)   &  1.03  &  0.049  \nl
\enddata
\tablenotetext{}{WRS -- Without Ringing Suppression}
\end{deluxetable}

\clearpage
\begin{deluxetable}{lclcl}
\tablecaption{Surface Brightness Test \label{tbl:bright}}
\tablewidth{0pt}
\tablehead{
& \multicolumn{2}{c}{Iteration 1} & \multicolumn{2}{c}{Iteration
20} \nl
\colhead{Image} & \colhead{log(MIGA/ISSA)} & \colhead{St. Dev.} &
\colhead{log(MIGA/ISSA)} & \colhead{St. Dev.} 
}
\startdata
12 $\mu$m Band  & & & & \nl
\hspace*{0.5cm} g075+02\tablenotemark{1}   & 0.00285  &  0.0193 & $\phantom{-}0.00021$  &  0.03913 \nl
\hspace*{0.5cm} g086+05   & 0.00299 &  0.0165 & $\phantom{-}0.00054$  &  0.04169 \nl
\hspace*{0.5cm} g102$-$03 & 0.00321 &  0.0310 &           $-0.00465$  &  0.08860 \nl
\hspace*{0.5cm} g120$-$01 & 0.00305 &  0.0232 &           $-0.00190$  &  0.05687 \nl
\hspace*{0.5cm} g141$-$02 & 0.00373 &  0.0453 &           $-0.00381$  &  0.08510 \nl
25 $\mu$m Band  & & & & \nl
\hspace*{0.5cm} g075+02 & 0.00165 &  0.0196 &             $-0.00077$  &  0.03201 \nl
\hspace*{0.5cm} g086+05 & 0.00113 &  0.0067 &   $\phantom{-}0.00055$  &  0.01750
 \nl
\hspace*{0.5cm} g102$-$03 & 0.00043 &  0.0067 &           $-0.00010$  &  0.02003
 \nl
\hspace*{0.5cm} g120$-$01 & 0.00058 &  0.0099 &           $-0.00044$  &  0.02386 \nl
\hspace*{0.5cm} g141$-$02 & 0.00183 &  0.0286 &           $-0.00228$  &  0.05743 \nl 
\enddata
\tablenotetext{1}{See Figure~\ref{fig:bright}}
\end{deluxetable}

\clearpage
\begin{deluxetable}{lllllllll}
\footnotesize
\tablecaption{Position Test Results \label{tbl:pos}}
\tablewidth{0pt}
\tablehead{ & \multicolumn{2}{c}{PSC Position} & \multicolumn{2}{c}{MIGA 12
$\mu$m} & \multicolumn{2}{c}{MIGA 25 $\mu$m} &
\multicolumn{2}{c}{Distance\tablenotemark{\dagger} ($''$)}  \nl
\colhead{Source} & \colhead{$l$} & \colhead{$b$} &
\colhead{$l$} & \colhead{$b$} & \colhead{$l$}
& \colhead{$b$} & \colhead{12 $\mu$m} & \colhead{25 $\mu$m}  
}
\startdata
$ 00067+6340 $ & $ 118.3409 $ & $ \phantom{-}1.4576 $ & $ 118.3385 $ & $ \phantom{-}1.4589 $ & $ 118.3388 $ & $ \phantom{-}1.4592 $ & $ \phantom{1}9.8 $ & $ \phantom{1}9.4 $ \nl
$ 00186+5940 $ & $ 119.1762 $ & $ -2.6994 $ & $ 119.1747 $ & $ -2.6980
$ & $ 119.1747 $ & $ -2.6991 $ & $ \phantom{1}7.5 $ & $ \phantom{1}5.5  $ \nl
$ 00340+6251 $ & $ 121.3003 $ & $ \phantom{-}0.3074 $ & $ 121.2981 $ & $ \phantom{-}0.3073 $ & $ 121.3002 $ & $ \phantom{-}0.3068 $ & $ \phantom{1}7.9 $ & $ \phantom{1}2.1 $ \nl
$ 00362+5924 $ & $ 121.3715 $ & $ -3.1555 $ & $ 121.3704 $ & $ -3.1534 $ & $ 121.3690 $ & $ -3.1559 $ & $ \phantom{1}8.7 $ & $ \phantom{1}9.2 $ \nl
$ 01071+6551 $ & $ 124.8524 $ & $ \phantom{-}3.3233 $ & $ 124.8476 $ & $ \phantom{-}3.3210 $ & $ 124.8492 $ & $ \phantom{-}3.3244 $ & $ 19.1 $ & $ 12.1 $ \nl
$ 01142+6306 $ & $ 125.8542 $ & $ \phantom{-}0.6440 $ & $ 125.8532 $ & $ \phantom{-}0.6449 $ & $ 125.8546 $ & $ \phantom{-}0.6457 $ & $ \phantom{1}4.8 $ & $ \phantom{1}6.3 $ \nl
$ 01145+5902 $ & $ 126.2820 $ & $ -3.3949 $ & $ 126.2779 $ & $ -3.3938 $ & $ 126.2792 $ & $ -3.3930 $ & $ 15.3 $ & $ 12.1 $ \nl
$ 01217+6049 $ & $ 126.9839 $ & $ -1.5298 $ & $ 126.9852 $ & $ -1.5254 $ & $ 126.9828 $ & $ -1.5302 $ & $ 16.5 $ & $ \phantom{1}4.3 $ \nl
$ 01261+6446 $ & $ 126.9464 $ & $ \phantom{-}2.4612 $ & $ 126.9441 $ & $ \phantom{-}2.4625 $ & $ 126.9438 $ & $ \phantom{-}2.4614 $ & $ \phantom{1}9.4 $ & $ \phantom{1}9.5 $ \nl
$ 01364+6038 $ & $ 128.7872 $ & $ -1.4185 $ & $ 128.7854 $ & $ -1.4179 $ & $ 128.7853 $ & $ -1.4181 $ & $ \phantom{1}6.8 $ & $ \phantom{1}6.8 $ \nl
$ 01443+6417 $ & $ 128.9559 $ & $ \phantom{-}2.3395 $ & $ 128.9548 $ & $ \phantom{-}2.3420 $ & $ 128.9536 $ & $ \phantom{-}2.3393 $ & $ \phantom{1}9.8 $ & $\phantom{1} 8.3 $ \nl
$ 01572+5844 $ & $ 131.7681 $ & $ -2.6935 $ & $ 131.7683 $ & $ -2.6912 $ & $ 131.7693 $ & $ -2.6929 $ & $ \phantom{1}8.2 $ & $ \phantom{1}4.8 $ \nl
$ 01577+6354 $ & $ 130.4745 $ & $ \phantom{-}2.3018 $ & $ 130.4735 $ & $ \phantom{-}2.3016 $ & $ 130.4735 $ & $ \phantom{-}2.3009 $ & $ \phantom{1}3.7 $ & $ \phantom{1}5.0 $ \nl
$ 02044+6031 $ & $ 132.1572 $ & $ -0.7257 $ & $ 132.1547 $ & $ -0.7247 $ & $ 132.1546 $ & $ -0.7247 $ & $ \phantom{1}9.0 $ & $ \phantom{1}9.9 $ \nl
$ 02086+6355 $ & $ 131.6201 $ & $ \phantom{-}2.6685 $ & $ 131.6179 $ & $ \phantom{-}2.6697 $ & $ 131.6190 $ & $ \phantom{-}2.6700 $ & $ \phantom{1}9.0 $ & $ \phantom{1}6.8 $ \nl
$ 02217+5712 $ & $ 135.3331 $ & $ -3.1573 $ & $ 135.3341 $ & $ -3.1529 $ & $ 135.3346 $ & $ -3.1542 $ & $ 16.3 $ & $ 12.4 $ \nl
$ 02347+5649 $ & $ 137.1203 $ & $ -2.8495 $ & $ 137.1191 $ & $ -2.8463 $ & $ 137.1183 $ & $ -2.8496 $ & $ 12.0 $ & $ \phantom{1}7.1 $ \nl
$ 02469+5646 $ & $ 138.6543 $ & $ -2.2057 $ & $ 138.6546 $ & $ -2.2044 $ & $ 138.6530 $ & $ -2.2062 $ & $ \phantom{1}4.7 $ & $ \phantom{1}5.0 $ \nl
$ 02473+5738 $ & $ 138.3201 $ & $ -1.3976 $ & $ 138.3203 $ & $ -1.3972 $ & $ 138.3199 $ & $ -1.3979 $ & $ \phantom{1}1.5 $ & $\phantom{1} 1.4 $ \nl
$ 03042+5850 $ & $ 139.7360 $ & $ \phantom{-}0.6994 $ & $ 139.7346 $ & $ \phantom{-}0.6992 $ & $ 139.7353 $ & $ \phantom{-}0.6997 $ & $\phantom{1} 5.0 $ & $\phantom{1} 2.7 $ \nl
$ 03094+5530 $ & $ 142.0154 $ & $ -1.8251 $ & $ 142.0152 $ & $ -1.8255 $ & $ 142.0154 $ & $ -1.8242 $ & $ \phantom{1}1.7 $ & $\phantom{1} 3.0 $ \nl
$ 03301+5658 $ & $ 143.6405 $ & $ \phantom{-}0.9657 $ & $ 143.6404 $ & $ \phantom{-}0.9654 $ & $ 143.6407 $ & $ \phantom{-}0.9654 $ & $ \phantom{1}1.1 $ & $ \phantom{1}1.5 $ \nl
$ 03385+5927 $ & $ 143.0775 $ & $ \phantom{-}3.6207 $ & $ 143.0759 $ & $ \phantom{-}3.6185 $ & $ 143.0748 $ & $ \phantom{-}3.6183 $ & $ \phantom{1}9.8 $ & $ 13.1 $ \nl
$ 03419+5429 $ & $ 146.4398 $ & $ -0.0588 $ & $ 146.440 $ & $ -0.0599 $ & $ 146.4395 $ & $ -0.0600 $ & $ \phantom{1}4.1 $ & $ \phantom{1}4.5 $ \nl
$ 20026+4018 $ & $ \phantom{1}76.4321 $ & $ \phantom{-}4.8081 $ & $ \phantom{1}76.4300 $ & $ \phantom{-}4.8086 $ & $ \phantom{1}76.4287 $ & $ \phantom{-}4.8091 $ & $ \phantom{1}7.9 $ & $ 12.9 $ \nl
$ 20028+3910 $ & $ \phantom{1}75.4887 $ & $ \phantom{-}4.1759 $ & $ \phantom{1}75.4913 $ & $ \phantom{-}4.1728 $ & $ \phantom{1}75.4902 $ & $ \phantom{-}4.1753 $ & $ 14.4 $ & $ \phantom{1}6.0 $ \nl
$ 20282+3604 $ & $ \phantom{1}75.7746 $ & $ -1.7196 $ & $ \phantom{1}75.7742 $ & $ -1.7170 $ & $ \phantom{1}75.7753 $ & $ -1.7180 $ & $\phantom{1}9.5 $ & $\phantom{1}6.3 $ \nl
$ 20422+4644 $ & $ \phantom{1}85.8464 $ & $ \phantom{-}2.6680 $ & $ \phantom{1}85.8437 $ & $ \phantom{-}2.6695 $ & $ \phantom{1}85.8440 $ & $ \phantom{-}2.6699 $ & $ 11.3 $ & $ 11.0 $ \nl
$ 20499+4657 $ & $ \phantom{1}86.8478 $ & $ \phantom{-}1.7934 $ & $ \phantom{1}86.8484 $ & $ \phantom{-}1.7952 $ & $ \phantom{1}86.8488 $ & $ \phantom{-}1.7946 $ & $ \phantom{1}6.9 $ & $ \phantom{1}5.6 $ \nl
$ 20502+4709 $ & $ \phantom{1}87.0338 $ & $ \phantom{-}1.8859 $ & $ \phantom{1}87.0340 $ & $ \phantom{-}1.8871 $ & $ \phantom{1}87.0341 $ & $ \phantom{-}1.8863 $ & $ \phantom{1}4.3 $ & $ \phantom{1}2.0 $ \nl
$ 20549+5245 $ & $ \phantom{1}91.7978 $ & $ \phantom{-}4.9220 $ & $ \phantom{1}91.7979 $ & $ \phantom{-}4.9231 $ & $ \phantom{1}91.7982 $ & $ \phantom{-}4.9223 $ & $ \phantom{1}4.1 $ & $ \phantom{1}1.7 $ \nl
$ 21015+4859 $ & $ \phantom{1}89.6478 $ & $ \phantom{-}1.6478 $ & $ \phantom{1}89.6496 $ & $ \phantom{-}1.6493 $ & $ \phantom{1}89.6496 $ & $ \phantom{-}1.6493 $ & $ \phantom{1}8.6 $ & $ \phantom{1}8.6 $ \nl
$ 21122+4900 $ & $ \phantom{1}90.8629 $ & $ \phantom{-}0.3731 $ & $ \phantom{1}90.8645 $ & $ \phantom{-}0.3739 $ & $ \phantom{1}90.8646 $ & $ \phantom{-}0.3747 $ & $ \phantom{1}6.4 $ & $ \phantom{1}8.4 $ \nl
$ 21167+5502 $ & $ \phantom{1}95.6773 $ & $ \phantom{-}4.0997 $ & $ \phantom{1}95.6749 $ & $ \phantom{-}4.1001 $ & $ \phantom{1}95.6751 $ & $ \phantom{-}4.1000 $ & $ \phantom{1}8.6 $ & $ \phantom{1}8.1 $ \nl
$ 21223+5114 $ & $ \phantom{1}93.5840 $ & $ \phantom{-}0.8016 $ & $ \phantom{1}93.5831 $ & $ \phantom{-}0.8018 $ & $ \phantom{1}93.5840 $ & $ \phantom{-}0.8008 $ & $ \phantom{1}3.2 $ & $ \phantom{1}3.0 $ \nl
$ 212\phantom{1}32+5705 $ & $ \phantom{1}97.7603 $ & $ \phantom{-}4.9227 $ & $ \phantom{1}97.7609 $ & $ \phantom{-}4.9253 $ & $ \phantom{1}97.7593 $ & $ \phantom{-}4.9235 $ & $ \phantom{1}9.6 $ & $ \phantom{1}4.4 $ \nl
$ 21282+5050 $ & $ \phantom{1}93.9862 $ & $ -0.1185 $ & $ \phantom{1}93.9863 $ & $ -0.1172 $ & $ \phantom{1}93.9863 $ & $ -0.1171 $ & $ \phantom{1}4.6 $ & $ \phantom{1}5.1 $ \nl
$ 21444+5053 $ & $ \phantom{1}95.9339 $ & $ -1.7734 $ & $ \phantom{1}95.9345 $ & $ -1.7744 $ & $ \phantom{1}95.9345 $ & $ -1.7744 $ & $ \phantom{1}4.3 $ & $ \phantom{1}4.2 $ \nl
$ 21453+5959 $ & $ 101.8655 $ & $ \phantom{-} 5.1417 $ & $ 101.8635 $ & $ \phantom{-}5.1409 $ & $ 101.8645 $ & $ \phantom{-}5.1410 $ & $ \phantom{1}7.7 $ & $ \phantom{1}4.5 $ \nl
$ 21475+5211 $ & $ \phantom{1}97.1263 $ & $ -1.0757 $ & $ \phantom{1}97.1246 $ & $ -1.0742 $ & $ \phantom{1}97.1242 $ & $ -1.0748 $ & $\phantom{1}8.0 $ & $\phantom{1}8.2 $ \nl
$ 22122+5745 $ & $ 103.3018 $ & $ \phantom{-}1.2718 $ & $ 103.3008 $ & $ \phantom{-}1.2722 $ & $ 103.3008 $ & $ \phantom{-}1.2712 $ & $ \phantom{1}3.8 $ & $ \phantom{1}4.1 $ \nl
$ 22248+6058 $ & $ 106.3922 $ & $ \phantom{-}3.0933 $ & $ 106.3902 $ & $ \phantom{-}3.0953 $ & $ 106.3914 $ & $ \phantom{-}3.0958 $ & $ 10.3 $ & $ \phantom{1}9.4 $ \nl
$ 22303+5950 $ & $ 106.3932 $ & $ \phantom{-}1.7784 $ & $ 106.3917 $ & $ \phantom{-}1.7798 $ & $ 106.3931 $ & $ \phantom{-}1.7813 $ & $ \phantom{1}7.5 $ & $ 10.4 $ \nl
$ 22413+5929 $ & $ 107.4251 $ & $ \phantom{-}0.7972 $ & $ 107.4241 $ & $ \phantom{-}0.7970 $ & $ 107.4248 $ & $ \phantom{-}0.7958 $ & $ \phantom{1}3.5 $ & $ \phantom{1}5.3 $ \nl
$ 22471+5902 $ & $ 107.8860 $ & $ \phantom{-}0.0507 $ & $ 107.8863 $ & $ \phantom{-}0.0503 $ & $ 107.8851 $ & $ \phantom{-}0.0503 $ & $ \phantom{1}1.8 $ & $ \phantom{1}3.6 $ \nl
$ 22489+6359 $ & $ 110.2892 $ & $ \phantom{-}4.3798 $ & $ 110.2887 $ & $ \phantom{-}4.3810 $ & $ 110.2893 $ & $ \phantom{-}4.3815 $ & $ \phantom{1}4.7 $ & $ \phantom{1}6.3 $ \nl
$ 23000+5932 $ & $ 109.5835 $ & $ -0.1895 $ & $ 109.5848 $ & $ -0.1874 $ & $ 109.5842 $ & $ -0.1886 $ & $ \phantom{1}8.8 $ & $ \phantom{1}4.3 $ \nl
$ 23106+6340 $ & $ 112.3522 $ & $ \phantom{-}3.1283 $ & $ 112.3498 $ & $ \phantom{-}3.1283 $ & $ 112.3503 $ & $ \phantom{-}3.1292 $ & $ \phantom{1}8.5 $ & $ \phantom{1}7.5 $ \nl
$ 23239+5754 $ & $ 111.8786 $ & $ -2.8500 $ & $ 111.8772 $ & $ -2.8486
$ & $ 111.8763 $ & $ -2.8496 $ & $ \phantom{1}7.1 $ & $ \phantom{1}8.4 $ \nl
$ 23281+5742 $ & $ 112.3421 $ & $ -3.2212 $ & $ 112.3382 $ & $ -3.2207 $ & $ 112.3401 $ & $ -3.2186 $ & $ 14.2 $ & $ 11.6 $ \nl
$ 23321+6545 $ & $ 115.2069 $ & $ \phantom{-}4.3214 $ & $ 115.2051 $ & $ \phantom{-}4.3203 $ & $ 115.2052 $ & $ \phantom{-}4.3221 $ & $ \phantom{1}7.7 $ & $ \phantom{1}6.6 $ \nl
$ 23592+6228 $ & $ 117.2816 $ & $ \phantom{-}0.4239 $ & $ 117.2787 $ & $ \phantom{-}0.4240 $ & $ 117.2799 $ & $ \phantom{-}0.4252 $ & $ 10.5 $ & $ \phantom{1}7.7 $ \nl
\enddata
\tablenotetext{}{NOTE---Galactic latitude and longitude ($l,b$) are given in degrees.}
\tablenotetext{\dagger}{Absolute offset between MIGA and PSC position at each wavelength}
\end{deluxetable}

\clearpage
\begin{deluxetable}{lllllllll}
\footnotesize
\tablecaption{Position Test -- 15$''$ Pixels\tablenotemark{1} \label{tbl:pos15}}
\tablewidth{0pt}
\tablehead{ & \multicolumn{2}{c}{PSC Position} & \multicolumn{2}{c}{ 12
$\mu$m} & \multicolumn{2}{c}{25 $\mu$m} &
\multicolumn{2}{c}{Distance ($''$)}  \nl
\colhead{Source} & \colhead{$l$} & \colhead{$b$} &
\colhead{$l$} & \colhead{$b$} & \colhead{$l$}
& \colhead{$b$} & \colhead{12 $\mu$m} & \colhead{25 $\mu$m}  
}
\startdata
02570+5844 & 138.9591 & \phantom{$-$}0.1584 & 138.9584 & \phantom{$-$}0.1577 & 138.9584 & \phantom{$-$}0.1576 & \phantom{1}4.7  & 5.0 \nl
02568+5931 & 138.5898 & \phantom{$-$}0.8357 & 138.5795 & \phantom{$-$}0.8379 & 138.5791 & \phantom{$-$}0.8375 & \phantom{1}8.1  &  7.2 \nl
03012+5942 & 138.9766 & \phantom{$-$}1.2617 & 138.9757 & \phantom{$-$}1.2628 & 138.9753 & \phantom{$-$}1.2625 &  \phantom{1}4.4  & 6.5 \nl
02552+5937 & 138.3453 & \phantom{$-$}0.8277 & 138.3444 & \phantom{$-$}0.829900 & 138.3451 & \phantom{$-$}0.8293 & \phantom{1}9.2  &  6.0 \nl
02504+5917 & 137.9660 & \phantom{$-$}0.2526 & 137.9648 & \phantom{$-$}0.2542 & 137.9653 & \phantom{$-$}0.2544 & \phantom{1}6.7  &  7.6 \nl
02474+5901 & 137.7270 & $-$0.1555 & 137.7271 & $-$0.1526 & 137.7254 & $-$0.1543 &   10.4  & 8.4 \nl
02515+6001 & 137.7552 & \phantom{$-$}0.9805 & 137.7538 & \phantom{$-$}0.9794 & 137.7543 & \phantom{$-$}0.9801 & \phantom{1}5.8  & 4.6 \nl
02401+5923 & 136.7289 & $-$0.2174 & 136.7281 & $-$0.2167 & 136.7290 & $-$0.2165 &   \phantom{1}4.1 & 3.2 \nl
02368+5955 & 136.1374 & \phantom{$-$}0.0901 & 136.1370 & \phantom{$-$}0.0912 & 136.1369 & \phantom{$-$}0.0913 & \phantom{1}4.1  &  4.7 \nl
02314+5942 & 135.5906 & $-$0.3742 & 135.5890 & $-$0.3741 & 135.5914 & $-$0.3743 &   \phantom{1}5.8  &  1.4 \nl
02355+6029 & 135.7592 & \phantom{$-$}0.5560 & 135.7581 & \phantom{$-$}0.5528 & 135.7581 & \phantom{$-$}0.5538 & 12.4  &  8.9  \nl
\enddata
\tablenotetext{1}{Regular MIGA pixel size}
\end{deluxetable}

\clearpage
\begin{deluxetable}{lllllllll}
\footnotesize
\tablecaption{Position Test -- 18$''$ Pixels\tablenotemark{1} \label{tbl:pos18}}
\tablewidth{0pt}
\tablehead{ & \multicolumn{2}{c}{PSC Position} & \multicolumn{2}{c}{ 12
$\mu$m} & \multicolumn{2}{c}{25 $\mu$m} &
\multicolumn{2}{c}{Distance ($''$)}  \nl
\colhead{Source} & \colhead{$l$} & \colhead{$b$} &
\colhead{$l$} & \colhead{$b$} & \colhead{$l$}
& \colhead{$b$} & \colhead{12 $\mu$m} & \colhead{25 $\mu$m}  
}
\startdata
02570+5844 & 138.9591 & \phantom{$-$}0.1584  & 138.9599 &  \phantom{$-$}0.1592 & 138.9600 &   \phantom{$-$}0.1592  &  \phantom{1}4.3  & \phantom{1}4.3 \nl
02568+5931 & 138.5798 & \phantom{$-$}0.8357  & 138.5803 &  \phantom{$-$}0.8355 & 138.5799 &   \phantom{$-$}0.8351  &  \phantom{1}0.9  & \phantom{1}2.3 \nl
03012+5942 & 138.9766 & \phantom{$-$}1.2617  & 138.9756 &  \phantom{$-$}1.2604  & 138.9754 &   \phantom{$-$}1.2600  &  \phantom{1}5.2  & \phantom{1}8.3 \nl
02552+5937 & 138.3453 & \phantom{$-$}0.8277  & 138.3436 &  \phantom{$-$}0.8307 & 138.3443 &   \phantom{$-$}0.8302  &  11.6  & 10.0 \nl
02504+5917 & 137.9660 & \phantom{$-$}0.2526  & 137.9633 &  \phantom{$-$}0.2550 & 137.9636 &   \phantom{$-$}0.2553  &  13.7  & 12.0 \nl
02474+5901 & 137.7270 & $-$0.1555 & 137.7272 & $-$0.15335 & 137.725 &  $-$0.1551  & \phantom{1}7.9  & \phantom{1}7.3 \nl
02515+6001 & 137.7552 & \phantom{$-$}0.9805  & 137.7552 &  \phantom{$-$}0.9802 & 137.7554 &   \phantom{$-$}0.9809  & \phantom{1}1.4  & \phantom{1}1.8 \nl
02401+5923 & 136.7289 & $-$0.2174 & 136.7287 & $-$0.2153 & 136.7298 &  $-$0.2150  & \phantom{1}7.7  & \phantom{1}9.6 \nl
02368+5955 & 136.1374 & \phantom{$-$}0.0901  & 136.1349 &  \phantom{$-$}0.0898 & 136.1345 &   \phantom{$-$}0.0898  & \phantom{1}8.7  & \phantom{1}8.7 \nl
02314+5942 & 135.5906 & $-$0.3742 & 135.5875 & $-$0.3740 & 135.5898 &  $-$0.3742  & \phantom{1}9.4  & \phantom{1}2.2 \nl
02355+6029 & 135.7592 & \phantom{$-$}0.5560  & 135.7596 &  \phantom{$-$}0.5535 & 135.7598 &  \phantom{$-$}0.5546  & \phantom{1}9.6  & \phantom{1}5.7 \nl
\enddata
\tablenotetext{1}{CGPS mosaic pixel size}
\end{deluxetable}

\clearpage
\begin{deluxetable}{lcccc}
\footnotesize
\tablecaption{Mosaic Test Results -- 12 $\mu$m \label{tbl:mosall12}}
\tablewidth{0pt}
\tablehead{
\colhead{} & \multicolumn{2}{c}{1$^{\mathrm st}$ Iteration}  & \multicolumn{2}{c}{20$^{\mathrm th}$ Iteration} \nl
\colhead{Plate} & \colhead{Mean Ratio $-$ 1 (\%)} &
\colhead{St. Dev. (\%)\tablenotemark{1}} & \colhead{Mean Ratio $-$ 1
(\%)} & \colhead{St. Dev. (\%)}
}
\startdata
p1572  &  $-$0.13 & 0.49  &  $-$0.07 &  3.19 \nl
p1573  &  $-$0.08 & 0.32  &  $-$0.13 &  2.00 \nl
p1574  &  \phantom{$-$}0.06 & 0.48  &  \phantom{$-$}0.09 &  1.50 \nl
p1575  &  \phantom{$-$}0.09 & 0.30  &  \phantom{$-$}0.13 &  1.78 \nl
Combined\tablenotemark{2} & $-$0.02  & 0.42  & 0.00  &   2.24 \nl
 \nl
Edges\tablenotemark{3} & $-$0.08 &  1.24  & $-$0.03  & 5.46 \nl
\enddata
\tablenotetext{1}{Standard deviation from pixel ratio values are reported as a measure of the scatter.}
\tablenotetext{2}{Total of 32294 pixels within the plates}
\tablenotetext{3}{Total of 4579 pixels along plate edges}
\end{deluxetable}

\clearpage

\begin{deluxetable}{lcccc}
\footnotesize
\tablecaption{Mosaic Test Results -- 25 $\mu$m \label{tbl:mosall25}}
\tablewidth{0pt}
\tablehead{
\colhead{} & \multicolumn{2}{c}{1$^{\mathrm st}$ Iteration}  & \multicolumn{2}{c}{20$^{\mathrm th}$ Iteration} \nl
\colhead{Plate} & \colhead{Mean Ratio $-$ 1 (\%)} &
\colhead{St. Dev. (\%)\tablenotemark{1}} & \colhead{Mean Ratio $-$ 1 (\%)} & \colhead{St. Dev. (\%)}
}
\startdata
p1572  &  $-$0.07 & 0.22  & $-$0.09 & 1.05 \nl
p1573  &  $-$0.08& 0.25  &  $-$0.04 & 1.03 \nl
p1574  &  \phantom{$-$}0.01 & 0.29  &  $-$0.01 & 0.65 \nl
p1575  &  \phantom{$-$}0.06 & 0.14  &  \phantom{$-$}0.05 & 0.63 \nl
Combined\tablenotemark{2} & $-$0.02  & 0.24 & $-$0.02 & 0.87 \nl
 \nl
Edges\tablenotemark{3}    & $-$0.03  & 0.67 & $-$0.03 & 2.05 \nl
\enddata
\tablenotetext{1}{Standard deviation of pixel ratio values are reported as a measure of the scatter.}
\tablenotetext{2}{Total of 32294 pixels within the plates}
\tablenotetext{3}{Total of 4579 pixels along plate edges}
\end{deluxetable}

\clearpage

\begin{deluxetable}{ccccc}
\footnotesize
\tablecaption{MIGA and IGA Mosaic Test Comparison\tablenotemark{\dagger} \label{tbl:moscomp} }
\tablewidth{0pt}
\tablehead{\colhead{} & \multicolumn{2}{c}{Internal St. Dev. (\%)} & \multicolumn{2}{c}{Edge St. Dev. (\%) } \nl \colhead{Wavelength ($\mu$m)} & \colhead{Iteration 1} & \colhead{Iteration 20} & \colhead{Iteration 1} & \colhead{Iteration 20}
}
\startdata
100 &  0.08  &  0.23  &  0.18   &  0.46 \nl
60  &  0.14  &  0.51  &  0.52   &  1.5  \nl
25  &  0.24  &  0.87  &  0.67   &  2.1  \nl
12  &  0.42  &  2.2   &  1.2    &  5.5  \nl
\enddata
\tablenotetext{\dagger}{Different plates were used for the IGA and
MIGA tests}
\end{deluxetable}

\clearpage

\begin{deluxetable}{ccclcl}
\tablecaption{Ringing Suppression Algorithm -- Surface Brightness Test \label{tbl:bnbbright}}
\tablewidth{0pt}
\tablehead{
\colhead{$l$} & \colhead{$b$} & \colhead{log(MIGA/ISSA)} & \colhead{St. Dev.} &
\colhead{log(WRS\tablenotemark{\dagger} /ISSA)} & \colhead{St. Dev.} 
}
\startdata
142.2 & $-$1.52  &  $-$0.0015 &  0.034  &  $-$0.0018  &  0.034 \nl
142.4  & $-$1.95 &  \phantom{$-$}0.0001 &  0.034 &  $-$0.0001 &  0.034 \nl
142.2 &  $-$2.41 &  $-$0.0033 &  0.037 &  $-$0.0039 &  0.036 \nl
141.5 &  $-$2.43 &  $-$0.0026 &  0.035 &  $-$0.0024  &  0.034  \nl
141.5 &  $-$1.81 &  $-$0.0028 &  0.036 &  $-$0.0029  &  0.036  \nl
\enddata
\tablenotetext{}{NOTE -- Coordinates are for the center of 10$'$ radius apertures}
\tablenotetext{\dagger}{WRS -- No Ringing Suppression applied}
\end{deluxetable}

\clearpage

\begin{deluxetable}{cccccccc}
\footnotesize
\tablecaption{Depth of Point Source Ringing\label{tbl:psrr}}
\tablewidth{0pt}
\tablehead{
& & \multicolumn{2}{c}{Flux (Jy)} & & \multicolumn{2}{c}{Depth(MIGA)/Depth(WRS)\tablenotemark{\dagger}} \nl \cline{5-8}
\colhead{Source} & \colhead{PSC Number} & \colhead{12 $\mu$m } & \colhead{25 $\mu$m } & \colhead{12 $\mu$m Left} & \colhead{12 $\mu$m Right} & \colhead{25 $\mu$m Left} & \colhead{25 $\mu$m Right} 
 }
\startdata
1  &  02459$+$6049  & 16.66 & \phantom{1}5.16  & $0.74$ & $0.66$ & $0.58$ & $0.72$ \nl
2  &  02445$+$6042  & \phantom{1}5.85  & 10.94 & $0.69$ & $0.49$ & $0.60$ & $0.04$ \nl
3  &  02455$+$6130  & \phantom{1}5.40  & \phantom{1}1.68  & $0.33$ & $0.52$ & $0.44$ & $0.63$ \nl
4  &  02358$+$5928  & 11.27 & \phantom{1}4.90  &  $0.45$ & $0.29$ & $0.06$ & $0.33$ \nl
5  &  02368$+$5955  & \phantom{1}4.75  & \phantom{1}2.33  & $0.47$ & $0.45$ & $0.26$ & $0.57$ \nl
6  &  02399$+$6012  & \phantom{1}3.00  & \phantom{1}0.86  & $0.66$ & $0.43$ & $0.55$ & $0.43$ \nl
7  &  02391$+$5959  & \phantom{1}1.01  & \phantom{1}0.32  & $0.64$ & $0.49$ & $0.90$ & $0.42$ \nl
8  &  02474$+$5901  & \phantom{1}6.96  & \phantom{1}2.15  & $0.38$ & $0.03$ & $0.28$ & $0.29$ \nl
9  &  02504$+$5917  & \phantom{1}4.60  & \phantom{1}1.35  & $0.28$ & $0.88$ & $0.48$ & $0.80$ \nl
10 &  02401$+$5923  & \phantom{1}3.83  & \phantom{1}2.17  & $0.29$ & $0.55$ & $0.30$ & $0.32$ \nl
\enddata
\tablenotetext{\dagger}{Without ringing suppression}
\end{deluxetable}

\end{document}